\documentclass{aa}  
\usepackage{graphicx}
\usepackage{float}
\newcommand{\orcid}[1]{\href{https://orcid.org/#1}{\protect\includegraphics[width=8pt]{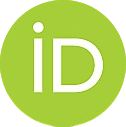}}}

\usepackage{txfonts}
\usepackage[colorlinks=true,urlcolor=Blue,citecolor=Blue,linkcolor=Blue,bookmarks=true]{hyperref}
\usepackage[dvipsnames]{xcolor}
\usepackage{lineno}

\begin{document} 

\title{Evolution of star clusters: Combined effects of initial substructures, primordial binary systems, and initial bulk rotation}
\titlerunning{Substructures, binaries, and rotation in star cluster evolution}
\authorrunning{Amiri et al.}
   
   \author{V. Amiri   \inst{1}\fnmsep\thanks{vahid.amiri@uni-heidelberg.de} \orcid{0009-0001-0633-9318}, 
   F. Flammini Dotti \inst{2,3,4}\fnmsep\thanks{ff2415@nyu.edu} \orcid{0000-0002-8881-3078},
   P. Berczik \inst{7}\orcid{0000-0003-4176-152X},
   X. Pang \inst{5,6}\orcid{0000-0003-3389-2263},
   M. Arca Sedda \inst{8,9,10}\orcid{0000-0002-3987-0519},
   \and
   R. Spurzem \inst{1,11,12}\fnmsep \orcid{0000-0003-2264-7203}
   }
   \institute{Astronomisches Rechen-Inst., Zentrum f\"ur Astronomie Univ. Heidelberg, Mönchhofstraße 12-14, 69120 Heidelberg, Germany
\and Department of Physics, New York University Abu Dhabi, PO Box 129188 Abu Dhabi, UAE
\and Center for Astrophysics and Space Science (CASS), New York University Abu Dhabi, PO Box 129188, Abu Dhabi, UAE
\and Dipartimento di Fisica, Sapienza, Universit\'a di Roma, P.le A.Moro 2, I-00185, Roma, Italy
\and
Department of Physics, Xi{'}an Jiaotong-Liverpool University, 111 Ren{'}ai Rd., Suzhou Dushu Lake Science and Education Innovation District, Suzhou Industrial Park, Suzhou 215123, P.R. China
\and
Shanghai Key Laboratory for Astrophysics, Shanghai Normal University, 
100 Guilin Road, Shanghai 200234, P. R. China
\and  
Main Astronomical Observatory, National Academy of Sciences of Ukraine, 27 Akademika Zabolotnoho St, 03143 Kyiv, Ukraine
\and
Gran Sasso Science Institute (GSSI), I-67100 Viale Francesco Crispi 7, L’Aquila (Italy)
\and
INFN, Laboratori Nazionali del Gran Sasso, I-67100 Assergi (Italy)
\and
INAF - Osservatorio Astronomico di Roma, I-00040 Monte Porzio Catone (Rome), Italy
\and
National Astronomical Observatories, Chinese Academy of Sciences, 20A Datun Rd., Chaoyang District, 100101, Beijing, China
\and
Kavli Institute for Astronomy and Astrophysics, Peking University, 5 Yi He Yuan Rd., Haidian Distri., Beijing 100871, P.R. China
   }

\date{}

\abstract
{Star clusters are born with internal substructures, which dissolve over a relatively short dynamical timescale. They also form with a significant population of primordial binary stars and a notable degree of initial bulk rotation. Improving observational data may offer deeper insight into how these initial conditions evolve with cluster age. Numerical simulations provide a powerful means to investigate how the interplay between these three fundamental properties shapes the long-term evolution of star clusters.}
{We aim to investigate how the interplay between initial substructure, bulk rotation, and primordial binaries shapes the dynamical evolution of star clusters, and to determine which of these initial conditions plays the dominant role in reproducing observational data.}
{We present a Fokker–Planck-based method to generate rotating, substructured initial conditions for star cluster simulations. We performed N-body simulations of star cluster dynamics using \textsc{Nbody6++GPU}, incorporating initial substructure, bulk rotation, and primordial binaries, while investigating the relative contributions of each process.  The results were qualitatively compared with observational data from the open cluster database of \citealt{pang22} and \citealt{pang23}.} 
{We find that stronger rotation slows substructure merging, delays mass segregation, and promotes disk- or bar-like shapes in inner regions. The interplay between rotation and substructure critically shapes early mass loss, with the high substructure suppressing early escapers in rotating clusters. The primordial binary fraction has only a minor effect on the survival of substructure, but a higher primordial binary fraction accelerates mass segregation in clusters with low initial substructure and contributes to enhanced late-time mass loss in rapidly rotating clusters. Notably, clusters with high initial rotation and substructure retain long-lived internal subclusters, providing insights into the early survival and dynamical evolution of hierarchical substructures.}
{}


   \maketitle
 \nolinenumbers

\section{Introduction}
\label{intro}

\par Gravitational interactions can be studied in stellar clusters (\citealt{heggiemillion, binney} and \citealt{Spurzem2023}) as they play an important role in galaxy formation. Nevertheless, modeling star clusters remains a challenging task, primarily because constructing realistic initial conditions is highly non-trivial. Recent observational and numerical studies have increasingly highlighted the importance of key properties, such as bulk rotation, primordial binary populations, and initial substructure, in determining the evolution of young open clusters.

\par According to observational data, substructures are frequently linked to young open clusters (\citealt{krause2020, pang22, greg2024} and \citealt{arnold2024}), and they merge during the early stages of star cluster formation (\citealt{kuhn}). Simulations indicate that stars are thought to form within molecular clouds in regions of high-density gas (\citealt{lada2003, kruijssen, lomax, ballone} and \citealt{chev}). As the gas begins to cool, gravitational forces progressively overcome the internal pressure, leading to the onset of star formation. During this contraction process, a network of filaments develops, which can channel and supply material into these star-forming regions (\citealt{kroupa01, zwart10, krause2020, karam, brooke2025} and \citealt{laverde2025}). Although these substructures vanish and generally become more centrally concentrated after several crossing times (see \citealt{binney} for the definition), many studies suggest that they can still significantly influence the long-term evolution of star clusters that originate through the hierarchical merging of such substructures (\citealt{goodwin, mc, alis,parker14, sills2018, ballone, daff, tor, greg2024, laverde2025, brooke2025} and \citealt{amiri}).

\par  Primordial binary systems play a crucial role in star clusters, acting as the main energy source that prevents runaway collapse and significantly influences the long-term evolution of star clusters. Although many simulations ignore primordial binary systems, observations reveal that the fraction of binaries in star clusters varies widely, typically between 0.2 and 0.5 (e.g., see \citealt{milone2012, li2013} and \citealt{ramirez2024}), and in some cases reaching even higher values (e.g., see \citealt{abt,duq,leinert93} and \citealt{offner23}). Observational studies also reported that the fraction of binary systems in open clusters is between 0.35 and 0.7 (e.g., see \citealt{sollimabinfractionOC}). In particular, some studies (e.g., \citealt{marks12} and \citealt{moe2017}) suggest that this fraction approaches unity for massive stars in clusters. Over the past decades, the impact of the primordial binaries on cluster evolution has been increasingly emphasized (e.g., \citealt{kroupa95, kroupa99, reip, kacz, wangdragon, belloni2, rastello, cour, dragonarca2, dragonarca3} and \citealt{cour2024}). 

\par Although initial rotation was historically neglected in studies of star cluster evolution, it has received increasing attention over the past two decades and has been incorporated into a growing number of theoretical and numerical studies. In star-forming regions, there is a degree of velocity gradient, according to observational studies (e.g., \citealt{ andre2007, rosolowsky2008, tobin2009, cottaar2015, tsitali2015, rigliaco2016} and \citealt{jadhav2024}). Observations show strong indications of rotation in young and globular clusters (e.g., \citealt{boberg2017, bellini2017, lanzoni2018b, ferraro2018, kamann2019, jindal2019, sollima2019, vasiliev2021} and \citealt{della2024p39}). Numerical studies (e.g., \citealt{spurzem99rot, kim2002, kim2004, fiestas2006fpasl, ernst2007,kim2008, hong2013, tiongco2017, livernois2021, livernois2022,kamlahrot,tiongco2022p8,scalco2023, pang2025} and \citealt{bianchini2026}) have demonstrated that cluster rotation plays a fundamental role in governing the dynamical evolution of stellar clusters. Rotation modifies the internal balance between gravity and centrifugal force, thereby influencing the rate of mass loss, the timing and depth of core collapse, and the overall structural evolution of the system. Although the exact cause of rotation is unknown, some research suggests that it may have been inherited from the progenitor cloud and as a result of subcluster mergers at the early stages of cluster formation (e.g., see \citealt{lee2016a, lee2016b, mapelli2017P1,lahen2020p3} and \citealt{ballone2021p19}), the host galaxy's tidal effect (e.g., see \citealt{vesperini2014, tiongco2017} and \citealt{tiongco2022p8}), or gas accretion during cluster development (e.g., see \citealt{chen2021}).

\par Thanks to recent observations and simulations, it has become increasingly important to study the evolution of star clusters from a more comprehensive perspective. Bulk rotation, primordial binary systems, and initial substructure are all expected to play fundamental roles in shaping the evolution of young open clusters; however, their combined influence remains poorly understood. This study sheds light on the evolution of young open clusters by simultaneously varying these key features, thereby providing a more realistic picture of star cluster evolution. In this paper, Sect.~\ref{ic} describes the initial conditions and the setup of the star clusters, Sect.~\ref{res} presents the results in detail, and Sect.~\ref{conc} provides the conclusions.

\section{Initial conditions }
\label{ic}

\begin{table}
\caption{Initial conditions for different models.}
\centering
\scalebox{0.92}{
    \begin{tabular}{|c||c|c|c|c|c|c|c|}

        \hline 
        & bf & $\omega$  & d & M$_{tot}$  & N & T$_{rh}$ & T$_{c}$ \\
         &  &  &  &  \tiny $[10^{3}M_\odot]$ & \tiny[$10^3$]  & \tiny[$10^2$Myr] & \tiny[Myr] \\
        \hline
        Model 1 & 0.1 & 0.2 & 0.25 & 5.827   & 11 & 7.0 & 7.0 \\
        \hline
        Model 2 & 0.1 & 0.2 & 2.0 & 5.827   & 11 & 6.7 & 10.8\\
        \hline
        Model 3 & 0.1 & 0.6 & 0.25 & 5.827   & 11 & 6.1 & 7.0\\
        \hline
        Model 4 & 0.1 & 0.6 & 2.0 & 5.827   & 11 & 7.3 & 9.6\\
        \hline
        Model 5 & 0.1 & 1.8 & 0.25 & 5.827   & 11 & 6.7 & 7.3\\
        \hline
        Model 6 & 0.1 & 1.8 & 2.0 & 5.827   & 11 & 6.7 & 8.7\\
        \hline
        Model 7 & 0.25 & 0.2 & 0.25 & 6.41   & 12.5 & 7.8 & 7.0\\
        \hline
        Model 8 & 0.25 & 0.2 & 2.0 & 6.41   & 12.5 & 7.8 & 9.0\\
        \hline
        Model 9 & 0.25 & 0.6 & 0.25 & 6.41   & 12.5 & 7.0 & 6.4\\
        \hline
        Model 10 & 0.25 & 0.6 & 2.0 & 6.41   & 12.5 & 7.5 & 9.0\\
        \hline
        Model 11 & 0.25 & 1.8 & 0.25 & 6.41   & 12.5 & 7.8 & 7.8\\
        \hline
        Model 12 & 0.25 & 1.8 & 2.0 & 6.41   & 12.5 & 7.0 & 6.3\\
        \hline
        
    \end{tabular}}

    \tablefoot{The columns give: the names of the models, the fraction of binary systems (bf), the rotation parameter ($\omega$), the degree of initial substructures (d), the total mass, the total number of particles, the half-mass relaxation time, and the crossing time, respectively.}
    \label{table}
\end{table}

\par For this study, we model the evolution of rotating star clusters using the state-of-the-art direct $N$-body code \textsc{Nbody6++GPU}, which is highly optimized for parallel computing and builds on the long development history of Aarseth’s direct integration codes (\citealt{Aarseth2003, Nitadori2012,wanggpu,kamlahmocca,Spurzem2023}). This code allows us to follow the detailed gravitational interactions of very large stellar systems, currently up to the order of a million bodies (\citealt{wangdragon,dragonarca2,dragonarca3,vergara2025,wu2026}). Setting up an initial star cluster that simultaneously includes substructures, bulk rotation, and a population of primordial binaries is a non-trivial task. Although \textsc{McLuster} (\citealt{kupper} and \citealt{mcluster}) is commonly used to generate initial conditions for star cluster simulations, this code does not provide a built-in option for introducing rotation. To construct our initial star cluster models, we adopt a new approach that allows us to incorporate initial bulk rotation, substructures, and primordial binaries into the cluster’s initial conditions.

\subsection{Implementation of initial substructures, primordial binaries and initial bulk rotation}

\par  In this study, we introduce an approach that incorporates all three ingredients into the initial configuration. Using this procedure, we generated and evolved twelve distinct cluster models. Our construction pipeline consists of three major stages. We begin by generating a rotating cluster following the scheme of \citealt{spurzem99rot}, which is a two-dimensional, orbit-averaged Fokker–Planck (FP) model that has been employed to investigate the dynamical evolution of rotating star clusters. This model expands the King model (\citealt{king}) for a flattened and rotating cluster model and numerically solves the FP equation to examine the effect of rotation in cluster evolution. The initially rotating model is characterized by parameters $W$ and $\omega$. We fix the dimensionless central potential $W_0=6.0$ (which is the standard parameter for the non-rotating King models) and use three values of $\omega$ (0.2, 0.6, and 1.8), where increasing $\omega$ means an increasing fraction of total energy is due to rotation (\citealt{spurzem99rot}). Hence, we study the effect of initial bulk rotation in low, moderate, and high rotation regimes. For each parameter combination, a discrete $N$-body model of one million particles is constructed ($N_r$). From this large rotating system, we then extract a subset of ten thousand stars using a selection algorithm designed to introduce a degree of initial substructure. The method proceeds by selecting several seed particles at random, including one that is forced to be the central star. Around these seeds, we define a radius $r_s$ that encloses $10^x$ particles, with 
$x=\log_{10}(N_r) \  d/d_{max}$,  where $d$ is the initial substructures degree, varying from $d_{min}=0$ (maximally substructured) to $d_{max}=3$ (no substructures). We compute $r=r_s(\frac{3-d}{2})^{3-d}$ and then randomly select a new particle from within this region. Each newly selected star becomes an additional seed, and the process is repeated until the sample contains 10000 particles. These particles retain their velocity vectors drawn from the smooth parent distribution. Because particles in dense subclumps inherit velocity dispersions characteristic of the global smooth potential, the subclumps are initially locally out of virial equilibrium but subject to global rotational velocity gradients and global velocity dispersion. In clusters with clumpy initial conditions, early dynamical processes occur on the local timescales of the substructures. The clumps interact and merge during violent relaxation, changing the cluster’s initial structure and kinematics (see the following sections). It is important to note that, in the absence of initial rotation, local violent relaxation reshapes the internal velocity distribution of each clump and leads to merging of substructures on a very short timescale (see e.g., \citealt{parker14, sills2018, laverde2025} and \citealt{amiri}).

\par After particle selection, to ensure equilibrium while preserving all structural features, we rescale the system by recalculating its kinetic and potential energies and applying appropriate scaling factors to the particle positions and velocities. This rescaling is performed considering the total cluster mass and ensures that the final configuration satisfies the virial equilibrium condition. Next, using \textsc{McLuster}, we generate a corresponding population of single and binary systems, and then map the binary properties (mass, separation, and relative velocity) and the single star masses onto the ten thousand selected particles from the rotating progenitor. For these simulations, binary fractions of 0.1 and 0.25 are considered to examine their effects (Note that a binary system contributes two stars to the sample; e.g., 9000 single stars plus 1000 binaries yield 11000 stars for bf = 0.1, as listed in Table  \ref{table}). Regarding the period distribution of primordial binaries, we follow \citealt{kroupa95} for stars with $M < 5 M_\odot$, and \citealt{sana} and \citealt{oh15} for $M > 5 M_\odot$. Binary pairing is performed following \citealt{sana}. Adding binaries causes small deviations in the kinetic and potential energies of the selected sample; therefore, we repeat the rescaling procedure once more. This process produces an initial cluster with the intended level of substructure and close to virial equilibrium.

\subsection{General parameters}
\par We adopt a stellar metallicity of 0.0005 following \citealt{kamlahrot, pang2026} and \citealt{amiri}, allowing direct comparison of the results across all works. We use a half-mass radius of $r_h =$ 5.0 pc, and an initial mass function (IMF) following \citealt{kroupa2001} with masses between 0.08 M$_\odot$ and 150.0 M$_\odot$. All clusters evolve within the standard tidal field of the solar neighborhood (\citealt{ninkovic}).

\par Regarding the primordial binary fraction, we aim to study the overall dynamical impact rather than to reproduce a specific observed value; however, the adopted primordial binary fractions for this study originate from observations. Gaia DR3-based studies report a total (soft+hard) binary fraction in open clusters ranging from $\sim$0.15 to 0.6 (\citealt{cordoni2023}), while \citealt{jiang2024} find a narrower range of 0.06–0.34 for mass ratios q $\geq$ 0.5, with a median of 0.17.  Given that \citealt{pang2026} show that most soft primordial binaries dissolve within 5-8 Myr, we adopt initial hard binary (maximum semi-major axis = 50 AU in our models) fractions of 0.1 and 0.25, bracketing this observed range while focusing on the surviving, dynamically relevant population. Note that although this cut-off excludes wide binaries, some low-mass binaries may still be dynamically soft, depending on their binding energies and the local velocity dispersion.

\par For the global rotation, we sample low, moderate, and high rotation regimes ($\omega_0 = 0.2,0.6,1.8$) to study the different degrees of initial rotation effects, rather than to reproduce a specific observed value. We combine low and high initial substructure with the mentioned range of $\omega$ and primordial binary fractions to obtain a comprehensive view of how these three features jointly shape cluster evolution.

\par We note that the global virial equilibrium adopted in our simulations is a simplified initial condition. Real embedded clusters may be out of equilibrium due to gas expulsion, stellar feedback, or hierarchical assembly (e.g., \citealt{lahen2020p3, rantala2024, della2024p39} and \citealt{lahen2025}). Our purely stellar-dynamical models allow us to isolate the combined effects of rotation, substructure, and primordial binaries without the additional effects of gas loss.

\par
\begin{figure}[htb]
   
\includegraphics[scale=0.7]{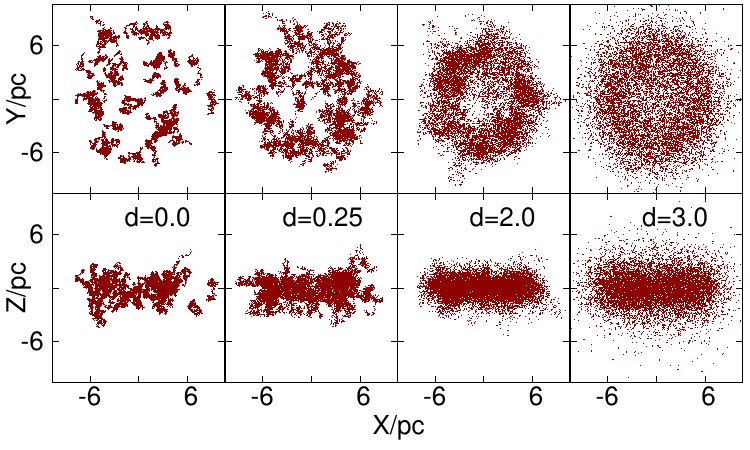}
 
    \caption{Initial distribution of stars in our models with the high degree of rotation and different levels of initial substructures, shown face-on in the top row and edge-on in the bottom row.}
    \label{inirot}
\end{figure}

 Figure~ \ref{inirot} shows the initial distribution of stars in models with the highest degree of rotation and different degrees of initial substructures in the X-Y and X–Z projections, confirming that the rotational flattening of the system (\citealt{bianchini2013}) is preserved throughout the selection procedure, while the substructures are well made based on the degree of the initial substructures. These conditions are identical to those used by \citealt{amiri}, allowing for reliable and meaningful comparisons.

\section{Results}
\label{res}
\par The main properties of our twelve models are listed in Table \ref{table} and the results of the simulations of the models are discussed in this section. We compared our results with the \citealt{pang22} and \citealt{pang23} open-cluster catalog. In this deprojected (3D) data sample, we retained clusters younger than 1 Gyr and excluded three clusters for which the reported core radius is negative. We sample 80 open clusters, whose ages range from 3.3 Myr to 955 Myr and whose current masses range from 42.5 to 2230 $M_\odot$. Further note that only those stars in the simulations that would be observable under the magnitude and mass limits of the \textit{Gaia} (\citealt{gaia} and \citealt{pang24}) are considered, while in other plots, these limitations are not taken into account. For this comparison, binary stars in the simulations are treated using their center-of-mass properties. 

Note that we performed one simulation for each of the twelve models. For the calculation of the $Q$ and $\lambda$ parameters (see Sect.~\ref{frsec} and Sect.~\ref{segsec}), we repeated the calculations using 20 random samples and adopted the average values. The shaded bands around the curves in the corresponding figures show the range of values obtained from these random samples.

\subsection{Evolution of substructures }
\label{frsec}
\begin{figure}[htb]
   \centering
\includegraphics[scale=0.38]{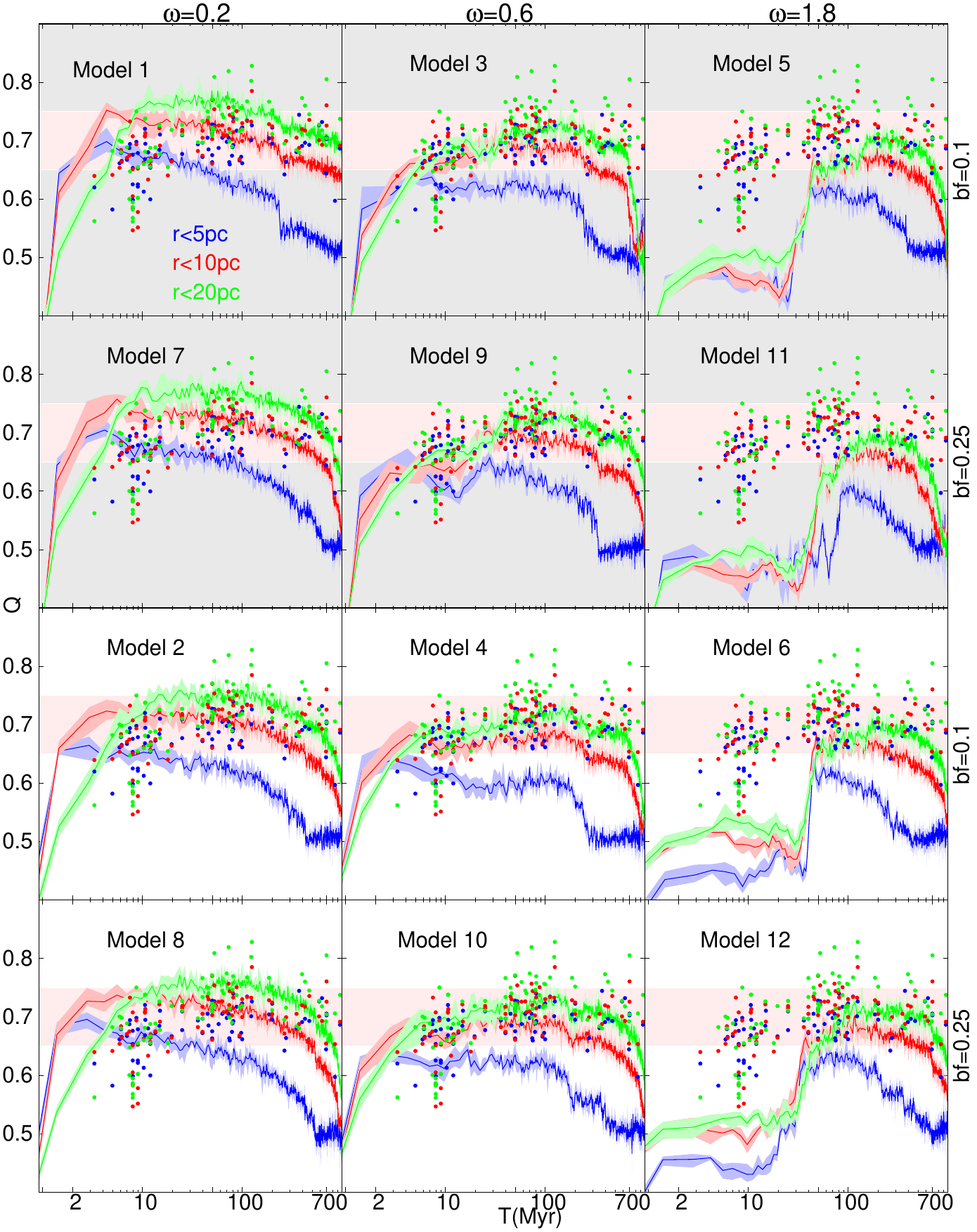}
 
    \caption{Evolution of initial substructures for all models in different regions (lines), compared with observational data (points) (\citealt{pang22} and \citealt{pang23}). The shaded range of 0.65–0.75 marks the regime where the particle distribution is approximately uniform. The shaded bands around the curves indicate the full range of possible Q values arising from randomization. Panel distribution: gray and white background panels show models with the high degree of initial substructures (d = 0.25) and the low degree of initial substructures (d = 2.0), respectively. Columns from left to right represent low ($\omega$ = 0.2), moderate ($\omega$ = 0.6), and high ($\omega$ = 1.8) rotation. Rows 1 and 3 show models with a primordial binary fraction of 10\%, while rows 2 and 4 show models with a primordial binary fraction of 25\%.}
    \label{fr}
\end{figure}

\par The evolution of substructures in a star cluster is of high importance, as it influences the entire dynamical history of the system. In substructured clusters, the local dynamical timescales within dense subregions are much shorter than the global crossing time of the cluster.  As a result, the dense areas tend to merge on a timescale of roughly one global crossing time (see the references in Sect.~\ref{intro}). However, observational studies (e.g., \citealt{pang22} and \citealt{pang23}) indicate that clusters with a high degree of initial substructures are still observed at ages of $\sim 15$ Myr, suggesting that the dissolution timescale of substructures can be significantly longer.

\par To study the lifetime of substructures in our models, Fig.~ \ref{fr} presents the evolution of the $Q$ parameter, which quantifies the initial substructures for the observable stars in all models, compared with observational data. The parameter is computed for sets of random stars within three different radii (r$<5$, r$<10$, and r$<20$ pc) following the approach of \citealt{goodwin}. The Q parameter is defined as Q $= l_{MST} / l_{cg}$, where $l_{cg}$ denotes the mean edge length of the complete $3$-D graph constructed from the stars in the region by connecting every star to all other stars, treating them as graph nodes. The quantity $l_{MST}$ corresponds to the mean edge length of the {\it minimum spanning tree} (MST)  formed by all stars in the region. Since these two quantities are fundamentally different, both measures must be normalized. Specifically, $l_{cg}$ is normalized by the radius of the analyzed region, whereas $l_{MST}$ is normalized by the characteristic length $(V/N_s)^{1/3}$, where $V$ is the volume of the region and $N_s$ is the number of sampled stars (\citealt{cart} and \citealt{ballone}). In the three-dimensional space, $Q$ values below $\sim0.7$ indicate the presence of substructure, values above $\sim0.7$ correspond to radially distributed systems, and values close to 0.7 represent approximately uniform stellar distributions (\citealt{ballone}). 

\par Figure~ \ref{fr} shows that the lifetime of the substructure depends strongly on the degree of rotation (initial degree of rotation increases from the left to the right columns of the figure). High initial rotation results in a significant centrifugal force that slows the collapse and merging of the substructures, thereby prolonging their survival. In models with the low rotation, even the high degree of initial substructures is erased within $\sim 2-5$ Myr within r<10 pc and r<20 pc, while in models with the moderate rotation, the disappearance time extends to $\sim 10-15$ Myr within those regions, and in models with the high degree of rotation, the substructure can persist for up to $\sim 70$ Myr. Our results indicate that a moderate degree of rotation plays a critical role in preserving initial substructures over longer timescales in young clusters, consistent with observational data. 

\par The $Q$ parameter also depends on the radius within which it is computed: larger volumes systematically yield higher $Q$ values, reflecting the smoother large-scale distribution of stars compared to the inner regions. For models with the high initial rotation, the inner region (r$<5$) retains substructures for almost the entire duration of the simulations. This occurs because centrifugal force prevents stars in the outer layers from efficiently migrating inward and mixing the substructures within the structures inside r$<5$. In addition, the lifetime of substructures in the inner region (r$<5$) exhibits two distinct phases in models with the low to moderate rotation. In this region, the substructures nearly vanish on a similar timescale as in the outer regions; however, the $Q$ parameter declines rapidly, which may mimic the signature of a highly substructured distribution. This happens because the core itself contracts dynamically: as mass segregation proceeds, massive stars sink toward the center, increasing the local density. This growing central concentration makes the MST more compact relative to the complete graph in the region, driving $Q$ downward even though the global substructure has already disappeared. In addition, the small number of stars within this region amplifies the sensitivity of $Q$ to these local density fluctuations.

\par It is of high importance to note that the sharp and nearly universal increase of the $Q$ parameter in the $\omega = 1.8$ models around $\sim 40$ Myr may be consistent with gravothermal-gravogyro evolution (see Sect.~\ref{lagrsec}) discussed in previous studies of rotation-driven contraction in star clusters (e.g., \citealt{hachisu1979, hachisu1982, spurzem99rot, kim2008} and \citealt{pang2025}). The initially low $Q$ values reflect the delayed merging of substructures due to centrifugal support, which stabilizes the system over several crossing times.

\par The strong drops in $Q$ observed at the final stages of simulations are likely numerical rather than physical, caused by a reduced number of stars remaining within the analyzed volume as escapers leave the cluster. It is also noteworthy that the primordial binary fraction has only a minor influence on the survival time of substructures, indicating that global rotational dynamics dominate binary interactions in determining the rate of substructure dissolution.

\subsection{Evolution of Lagrangian radii}
\label{lagrsec}

\begin{figure}[htb]
   \centering
\includegraphics[scale=0.42]{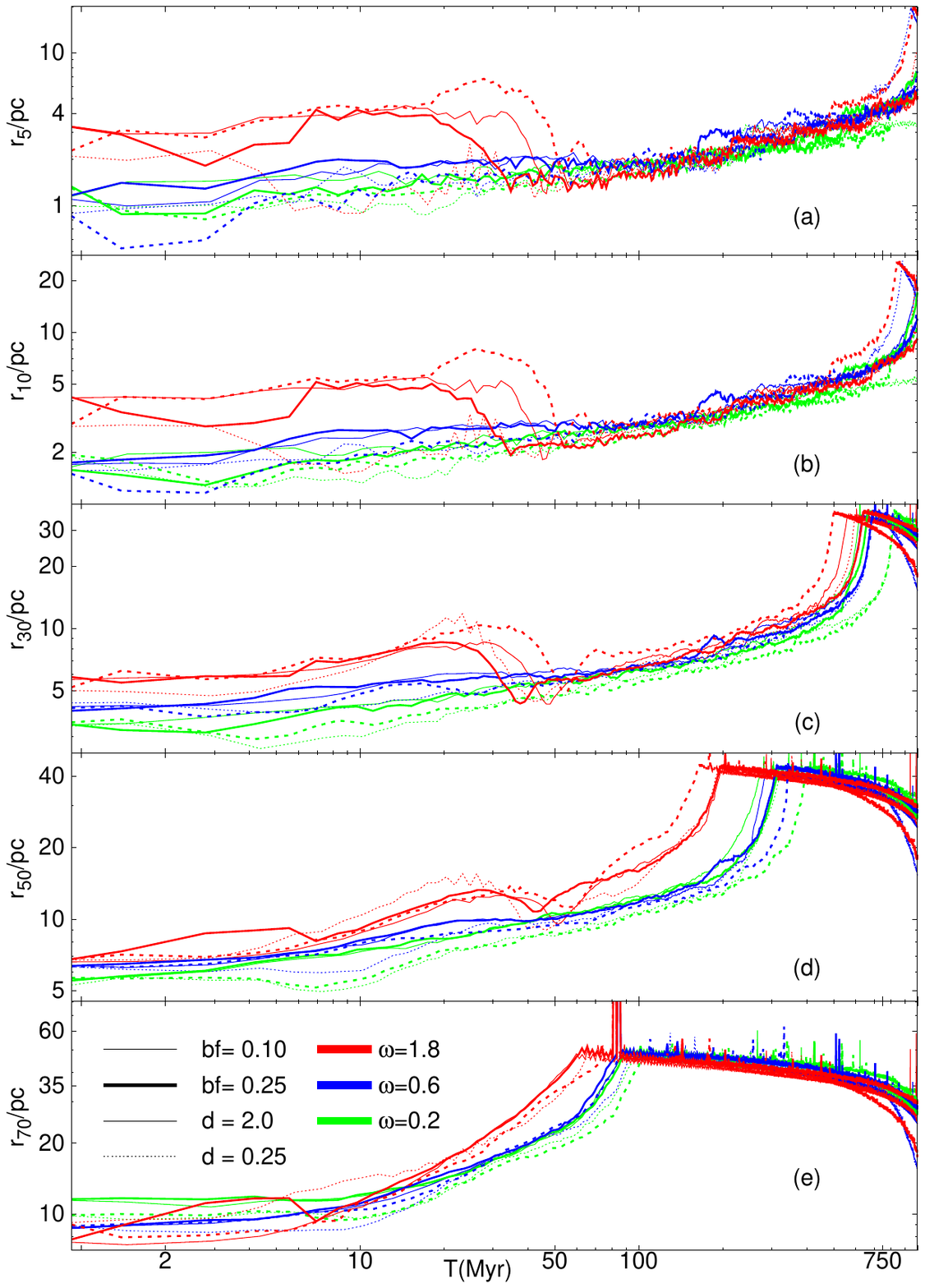}
 
    \caption{Evolution of different Lagrangian radii for all models. Here, dashed lines represent models with a high degree of initial substructure (d=0.25), while solid lines correspond to models with a low degree of initial substructure (d = 2.0). The green, blue, and red lines indicate models with the low ($\omega $ = 0.2), moderate ($\omega$ = 0.6), and high ($\omega$ = 1.8) initial rotation, respectively. Thin and thick lines denote models with 10\% and 25\% primordial binaries, respectively.}
    \label{lagr}
\end{figure}

\begin{figure}[htb]
   \centering
\includegraphics[scale=0.42]{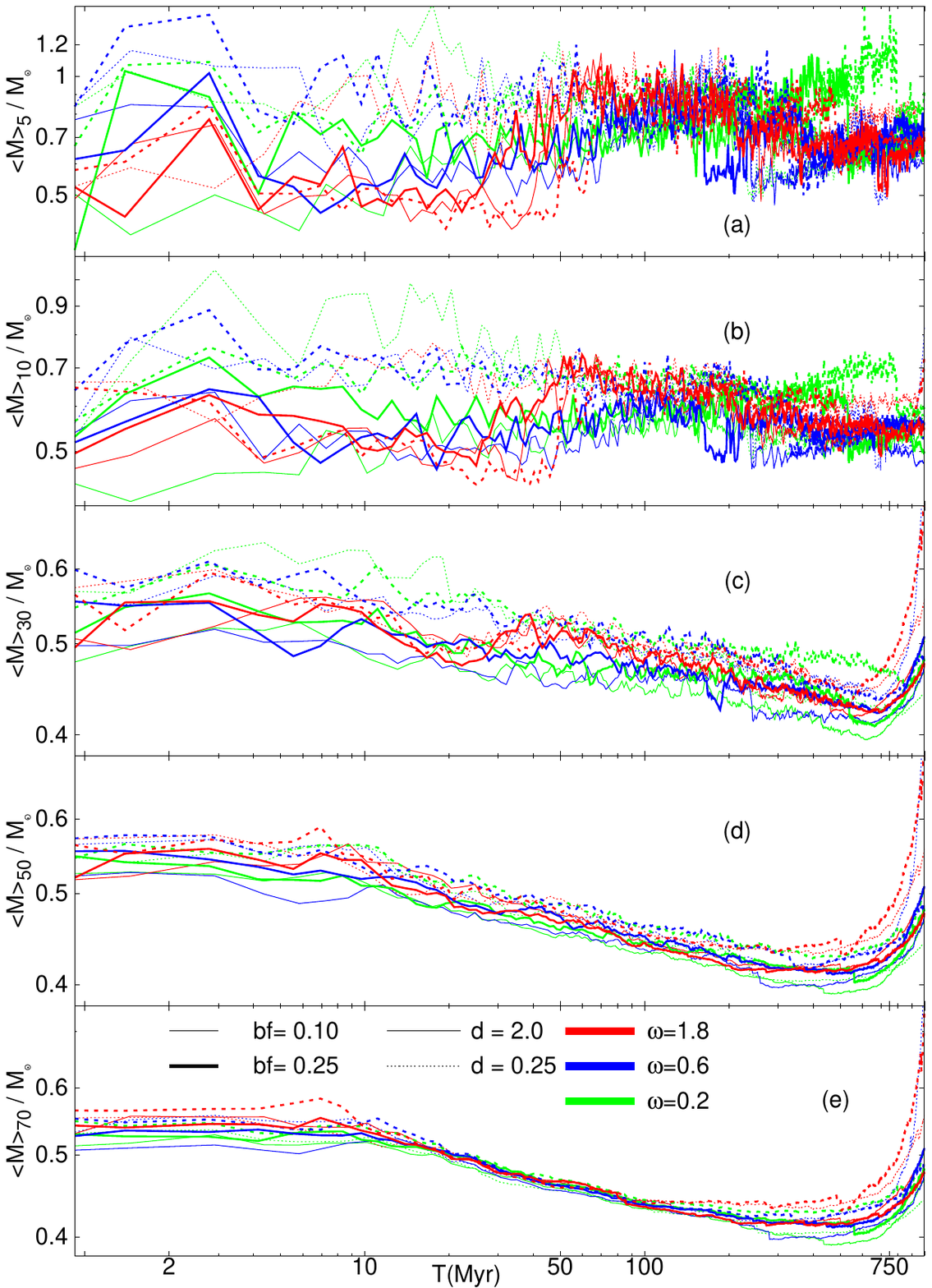}
 
    \caption{Evolution of average mass enclosed in different Lagrangian radii for all models. The models are shown using the same line styles and colors as in Fig.~ \ref{lagr}. }
    \label{lagrm}
\end{figure}

\par The global evolution of the models can be analyzed by examining how the Lagrangian radii change over time, along with the average mass contained within these radii. Figure~ \ref{lagr} shows the evolution of the different Lagrangian radii (with respect to the initial total mass) for all simulated models, while Fig.~ \ref{lagrm} displays the temporal evolution of the average mass enclosed within those radii. Together, these plots show how the interplay between initial rotation, initial substructures, and the binary fraction shapes the structural evolution of the clusters.

\par As Fig.~ \ref{lagr} shows, in the first $\sim 10$ Myr, clusters that begin with the low rotation and the high degree of initial substructures experience a noticeable contraction in $r_{10}$, $r_{30}$, and $r_{50}$, consistent with the results from \citealt{amiri}. A similar behavior is observed in the highly substructured model, with the low initial binary fraction and high degree of rotation showing a mild contraction in $r_{5}$ and $r_{10}$. During the early stage, the average stellar mass enclosed within the inner Lagrangian radii is systematically higher in models with stronger initial substructure and with the low to moderate degree of rotation. Recent observational studies (e.g., \citealt{pang22} and \citealt{della2024p39}) show that many embedded clusters younger than $\sim$30 Myr are expanding. However, in our purely dynamical setup, simulations do not show a universal expansion phase and indicate that the initial rotation and substructure level can strongly influence whether young clusters expand or contract. The high fraction of observed expanding young clusters may therefore reflect specific initial conditions or additional processes, such as residual gas expulsion, which is not included in our models.

\par In models with the high degree of initial rotation, a similar increase in the central average mass appears only after a delay, particularly when the fraction of primordial binaries is low. These trends can be understood through the dynamical timescales of the initial substructures. Substructures have much shorter crossing times than the whole cluster. Consequently, when the global rotation is low, the substructures collide and merge more efficiently. This merging accelerates the rate of close stellar encounters, allowing low-mass stars to gain kinetic energy and migrate toward the outer, more diffuse regions of the system. Conversely, the substructures enriched with the higher-mass stars lose energy and sink toward the center, producing a denser core (\citealt{mc}). This mechanism reflects a rapid onset of dynamical mass segregation (\citealt{alis}). Additionally, Fig.~ \ref{lagrm} reveals that, by $50$ Myr, all models with the higher initial substructures show consistently larger mean stellar masses within all radii, confirming that initial substructure substantially enhances mass segregation, consistent with the reported results by \citealt{alis} and \citealt{amiri}.

\par In agreement with studies (e.g., \citealt{spurzem99rot, kim2008} and \citealt{pang2025}) reporting that rotation leads to contraction in inner regions of the clusters, this study shows that, before $\sim 60$ Myr, all models with a high initial degree of rotation exhibit a pronounced contraction in the inner Lagrangian radii ($r_{5}$–$r_{50}$). The timing and strength of this contraction depend sensitively on the combined effects of the initial degree of substructures and the primordial binary fraction. Although the initial $T_{rh}$ is considerably longer than the epoch at which the contraction becomes prominent, the evolution of the inner regions is governed by much shorter local dynamical and relaxation times that develop in the dense central regions following the early evolution of the cluster.  This allows contraction of the inner regions within the first $\sim 60$ Myr, and even earlier than $\sim 10$ Myr for $r_5$ and $r_{10}$ in Model 5. This behavior is consistent with the onset of gravothermal-gravogyro evolution (\citealt{hachisu1979} and \citealt{hachisu1982}), in which the redistribution of angular momentum, coupled with gravothermal energy transport, triggers a rapid central contraction. In particular, when the initial rotation is high, a highly substructured initial configuration enhances early mass segregation and angular-momentum redistribution, while a low binary fraction reduces internal energy sources capable of delaying collapse. Together, these conditions may favor gravothermal-gravogyro evolution (for recent direct N-body models (without initial fractal structure) exhibiting gravothermal and gravogyro collapse, also see \citealt{kamlahrot}).

\par At later evolutionary stages, the structural evolution of the outer cluster regions diverges more clearly between models. Clusters with a high initial degree of rotation develop more extended outer envelopes, which subsequently reach a stable configuration (consistent with \citealt{pang2025}). In contrast, this behavior is substantially delayed in models with the low initial rotation. In these regimes, initial substructures further prolong this delay by promoting early local relaxation and energy dissipation in substructures, which slows the coherent global expansion of the outer regions.

\subsection{Changing in mass segregation }
\label{segsec}
\begin{figure}[htb]
   \centering
\includegraphics[scale=0.38]{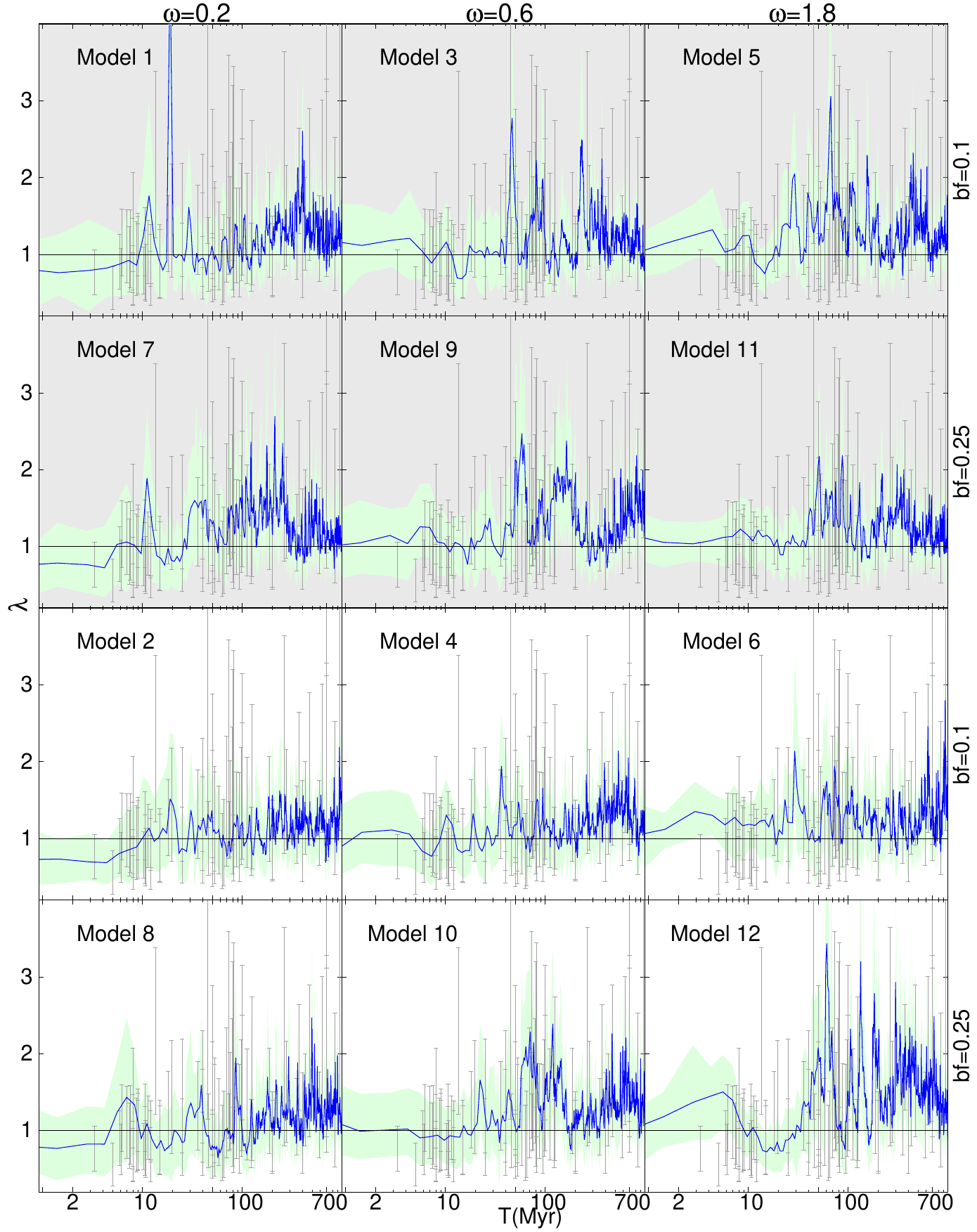}
 
    \caption{Evolution of mass segregation for all models (lines) compared to observational data (points with error bars)  (\citealt{pang22} and \citealt{pang23}). Green areas show all possible $\lambda$ values for the sampled stars (same as vertical error bars for observational data). Panel distribution represents different models, as in Fig.~ \ref{fr}.}
    \label{seg}
\end{figure}

\par Mass segregation refers to the tendency of massive stars to migrate toward the center of a cluster. This effect can be either primordial or dynamical in origin. Observations of very young star clusters already show signs of mass segregation, which is often classified as primordial (\citealt{deg} and \citealt{gou}). However, the origin of primordial mass segregation remains uncertain. Some studies suggest it may be related to competitive accretion during cluster formation (\citealt{bonnell2006}), the IMF (\citealt{bamg}), the formation of massive stars within substructures in their molecular cloud progenitors (\citealt{Allison10} and \citealt{dorval2015}), or it could arise entirely from dynamical processes operating on short timescales, such as the merging of substructures within the cluster (\citealt{alis, parker16} and \citealt{amiri}). To identify the degree of mass segregation, $\lambda$, in our models, we followed \citealt{mst} method, which compares the ratio of the MST length of a random sample to that of the top N most massive stars (NMST\footnote[1]{NMST = 10 in this study.}) to produce a quantitative measure of mass segregation. This quantifies the degree of mass segregation: values of $\lambda \approx 1$ indicate an absence of segregation, values below unity correspond to inverse segregation, and larger values represent progressively stronger mass segregation.

\par Figure~\ref{seg} shows the time evolution of $\lambda$ for all simulated models, restricted to observable stars, along with comparisons to the observational data. Previous studies have shown that the measured value of $\lambda$ depends sensitively on the choice of the number of selected top-most massive stars (NMST) (\citealt{Allison10, parker16, Hetem} and \citealt{parker22}). When NMST is large, the selected stellar sample spans a wider mass interval. As a result, many of the most massive stars included in the calculation have masses comparable to those drawn in the corresponding random samples. This reduces the contrast between the two distributions and drives the value of $\lambda$ toward unity. Accordingly, increasing NMST systematically shifts $\lambda$ closer to one, independent of the underlying dynamical state of the cluster. We clarify that the observational data show a broad range of $\lambda$ values over the considered age range. The simulations broadly reproduce the overall evolutionary trend of $\lambda$, although individual models do not necessarily match every observed cluster. This comparison is intended to assess the qualitative consistency of the simulated evolutionary behavior rather than to provide a unique fit to the observational sample.

\par  As shown in \citealt{amiri}, a high degree of initial substructures alone is sufficient to start mass segregation at early times. This occurs because dynamical mass segregation can be inherited by the final, smooth cluster once the substructures dissolve (\citealt{mc} and \citealt{livernois2021}) and can arise dynamically during the initial stages of star cluster evolution (\citealt{brooke2025}). However, when massive stars become concentrated in the central regions, their large encounter rates significantly alter this picture. Frequent strong interactions can eject massive stars from the core, or even from the cluster altogether. As a result, the initially strong mass segregation is transient and persists only for a short period of time (\citealt{brooke2025}).

\par Understanding the evolution of mass segregation becomes extremely challenging when initial substructures and rotation act together. Once the substructures have merged, the clusters progressively develop clearer and more persistent signatures of mass segregation. In this case, two processes dominate the early evolution: centrifugal forces induced by rotation and the rapid merging of initial substructures. Their combined effect drives an initially highly violent dynamical phase at the beginning of the simulation, which plays a crucial role in shaping the subsequent evolution of mass segregation. Global rotation has a consistently delaying effect on the emergence of strong mass segregation. The centrifugal barrier slows the merging of substructures (\citealt{livernois2021}) and inward drift of massive stars and reduces the rate at which subregions collapse toward the cluster center. This is especially evident in models with the high degree of initial substructure, where rotation prolongs the lifetime of dense substructures and prevents their mass content from efficiently sinking to the core. 

\par Two important features can be deduced from Fig.~ \ref{seg}: (i) low-rotation models ($\omega$ = 0.2) exhibit a short early phase of inverse mass segregation ($\lambda < 1$) during the first $\sim$5 Myr. This is likely related to the rapid merging of substructures, where strong dynamical interactions temporarily push massive stars outward before the system reaches a more relaxed state. The initial inverse segregation is later reversed again by two-body encounters and stellar evolution; (ii) high-rotation models with the low initial substructure (Models 6 and 12) show an early peak in $\lambda$ followed by a rapid decline, indicating enhanced mass segregation compared to other models. The early angular momentum redistribution in these models (see Sect.~\ref{momentumsec}) reduces the centrifugal support in the central regions, allowing massive stars to migrate inward. The resulting dense core then experiences strong stellar encounters that eject massive stars (\citealt{brooke2025}), leading to the subsequent decrease in $\lambda$.

\par The combined influence of initial substructures and primordial binary populations plays a decisive role in changing the degree of mass segregation (\citealt{kupper, parker11, parker14, parker16} and \citealt{parker22}). In models with the low binary fractions, strong initial substructures enhance mass segregation because locally segregated massive stars within subregions migrate to the cluster’s center during the merging of substructures (\citealt{mc}). In these cases, the merging process preserves the spatial distribution of massive stars inherited from the initial substructures, producing a clear central concentration after the substructures eventually dissolve. By contrast, in clusters with the low degree of initial substructures but the high fraction of initial binaries, binaries become important.  Gravitational interactions involving hard primordial binaries drive low-mass stars outward, while massive stars remain centrally concentrated, and this results in signals of mass segregation in a short time (\citealt{laverde2025}). Note that increasing the binary fraction from 0.10 to 0.25 only increases the total mass and number of stars by $\sim$10–14\%, and the corresponding change in $t_{rh}$ is minor (see Table \ref{table}). Therefore, we attribute the observed effect mainly to binary heating. In highly substructured models ($d = 0.25$), the early evolution is primarily governed by the merging of substructures.

\subsection{Mass loss and escapers}
\label{masssec}
\begin{figure}[htb]
   \centering
\includegraphics[scale=0.42]{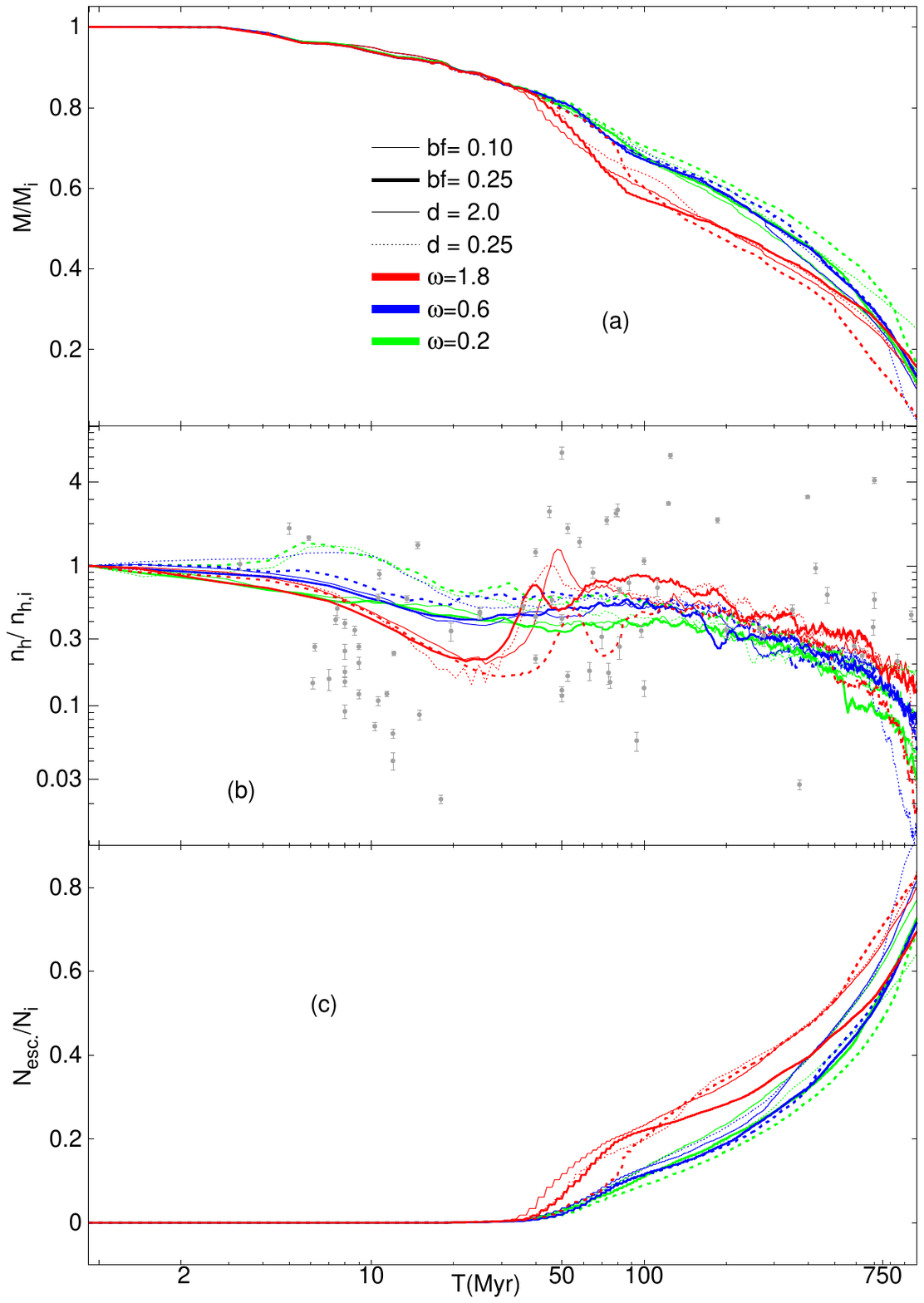}
 
    \caption{Evolution of mass (panel a), scaled number density in half-mass radius (lines)  compared to observational data (points) (\citealt{pang22} and \citealt{pang23}) in half-light radius (panel b), and the fraction of escapers (panel c) for all models. The models use the same line styles and colors as in Fig.~ \ref{lagr}.}
    \label{mass}
\end{figure}

\par In this section, we analyze the results for all models in terms of mass loss and the number of escapers in the presence of the standard model for the tidal field. \citealt{rossi} and \citealt{amiri} showed that initial substructures enhance early mass loss and the fraction of escapers. In the absence of rotation, highly substructured clusters undergo rapid local collapses within substructures. These collapses efficiently convert gravitational potential energy into random stellar motions, leading to the prompt ejection of stars during the first few tens of Myr. The resulting mass-loss is therefore dominated by the early dissolution of substructure.

\par To study the combined effects of primordial binaries, initial rotation, and initial substructures, we analyze Fig.~ \ref{mass}. Based on the panels (a) and (c) of Fig.~ \ref{mass}, however, rotation has a significant impact on the escapers and the mass loss of the cluster.  We note that during the first $\sim 30$ Myr, the decrease in the cluster mass for all models is primarily driven by stellar evolution.  After $\sim$40 Myr, models with larger initial rotation start to experience a higher mass loss compared to models with the low or moderate rotation. However, the impact of the initial substructure on mass loss is not monotonic with time: during the period between $\sim$40 and 60 Myr, a stronger initial substructure leads to a reduction in the mass loss. This time-dependent behavior reveals an important difference compared to the non-rotating case, where the presence of initial substructure enhances mass loss from the early stages of evolution (\citealt{amiri}). In the presence of initial rotation, the centrifugal force prolongs the lifetime of the substructure. This delays the onset of strong dynamical merging of substructures. Consequently, rotation becomes the primary regulator of mass loss in this regime: models with the higher rotation lose more mass (consistent with \citealt{spurzem99rot, kamlahrot} and \citealt{pang2025}). Part of the available kinetic energy is stored in ordered rotational motion rather than being transferred directly to random velocities. As a result, initially substructured rotating models experience milder early collapses (\citealt{livernois2021}) and reduced early encounter rates. We note that, in model 5, two results concern different parts of the cluster and different stages of its evolution: the early decrease of $r_5$ and $r_{10}$ (t$<$10 Myr) is a local response of the core. Angular momentum is transported away from the core, reducing its centrifugal support and causing it to contract. This does not mean that more stars immediately escape.  At the same time, the strong rotation and substructure in this model delay the merging of the substructures and therefore the conversion of ordered rotational motion into random stellar motions. The core can thus contract while the global dynamical mass loss remains low. These processes do not occur on the same timescale or affect the same regions of the cluster.

\par \citealt{pang2025} showed that increasing the initial binary fraction generally reduces the fraction of escaping stars. Our simulations, however, show that when initial substructures are included, the situation changes. In models with both initial substructures and initial rotation, clusters with a high initial binary fraction exhibit nearly the same escaper fraction as those with fewer binaries.

\par At later times, once the substructures have dissolved in outer regions (see Fig.~ \ref{fr} for r<10 pc and r<20 pc), the dominant dynamical channels shift. In highly rotating systems, differential rotation, shear, and large-scale torques begin to redistribute angular momentum and progressively convert ordered rotational energy into random stellar motions. This process is amplified when initial substructure and primordial binaries are present in the model . Substructure in inner regions ( see Fig.~ \ref{fr} for r<5 pc) enhances the frequency and strength of close encounters, while binaries act as internal energy sources (\citealt{binney}). Together, these mechanisms efficiently heat the cluster, raise stellar velocities, and enhance mass loss. After approximately 40 Myr (70 Myr for the model with the high fraction of primordial binaries and high degree of substructures), models with strong initial rotation exhibit significantly higher escape rates compared to other models. This effect is further amplified when high rotation is combined with a strong degree of initial substructures and a large primordial binary fraction.

\par This sequence resolves the apparent contradiction with \citealt{amiri}: initial substructures increase mass loss in non-rotating clusters because violent early collapses dominate the energy budget, but in rotating clusters, the degree of initial substructure produces milder early dynamics, suppressing early escapers. Thus, the mass-loss behavior of substructured clusters depends critically on the interplay between substructure and rotation and on which energy-conversion channel (violent collapse vs. rotational torques and binary heating) dominates at a given epoch.

\par Panel (b) shows the time evolution of the half-mass density ratio of observable stars, normalized to its initial value, for all models. In systems with strong initial rotation, centrifugal support delays central contraction until the initial substructure is progressively erased. Central density in the high rotation models first declines over the initial $\sim$5-30 Myr, since strong centrifugal support suppresses early cluster contraction while stellar-evolution mass loss further lowers the density. This decline is most pronounced in the highly substructured models, where rotation additionally prolongs substructure survival. Once these substructures have largely dissolved, the subsequent central density increase is consistent with gravothermal-gravogyro contraction. During this phase, the scaled number density remains broadly consistent with observational constraints for all models up to $\sim 700$ Myr. We note that at early times (< 10 Myr), some observed clusters show density ratios lower than those in our models. This likely reflects differences in initial conditions for real clusters (primordial concentration, gas expulsion, or extinction-driven incompleteness). Beyond this time, the density ratio in all models drops significantly below the observed values, primarily because a large fraction of particles escapes from the clusters. Investigating this ratio at later evolutionary stages, therefore, requires simulations with a larger number of stars, ensuring that a sufficient population remains gravitationally bound.

\subsection{Evolution of primordial binary fraction}
\label{binsec}
\begin{figure}[htb]
   \centering
\includegraphics[scale=0.42]{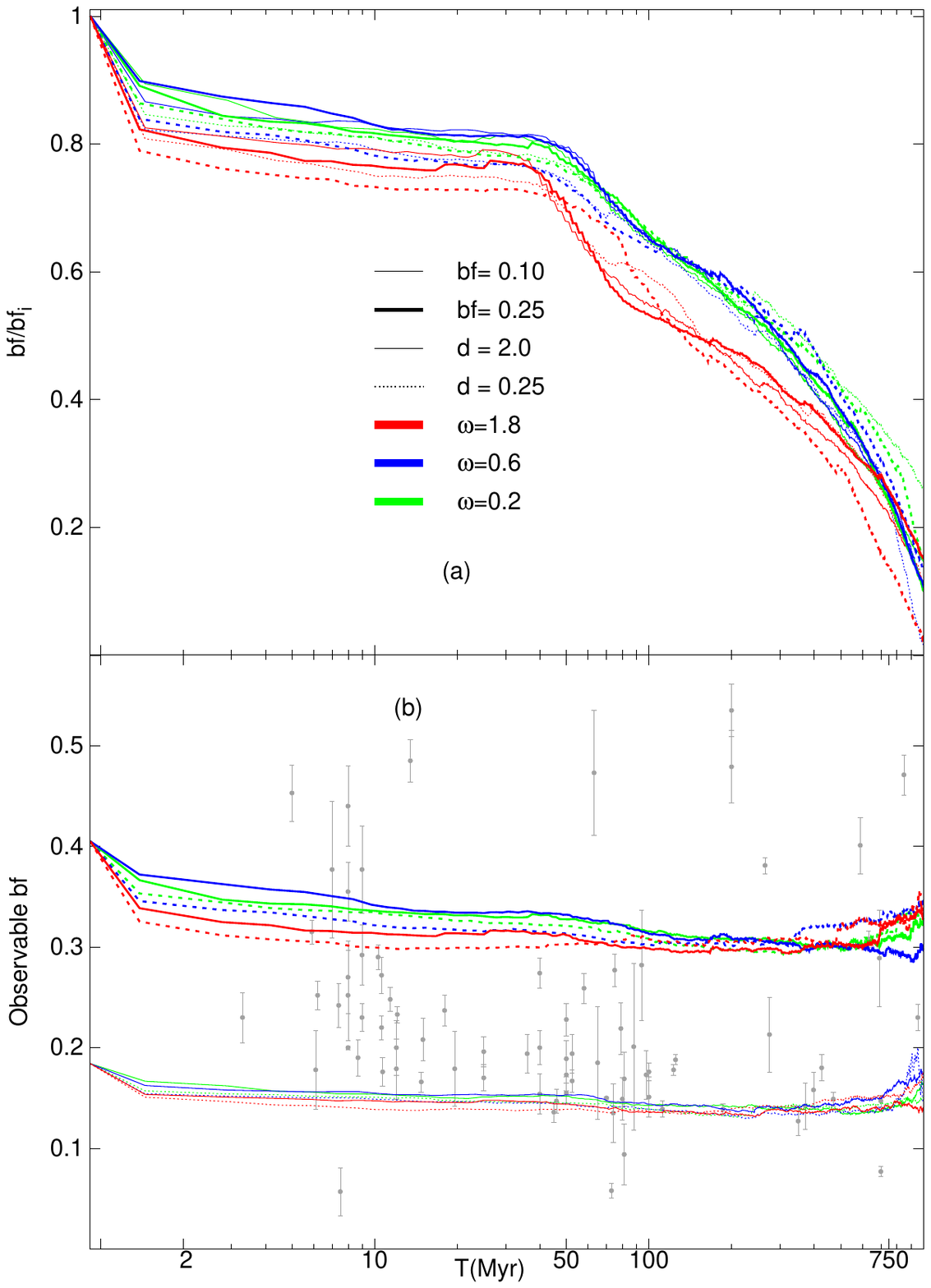}
 
    \caption{Evolution of the fraction of binary systems (panel a) and the observable fraction of binaries (lines) compared to the observational data (points) (\citealt{pang22} and \citealt{pang23}) (panel b) for all models. The models are shown using the same line styles and colors as in Fig.~ \ref{lagr}.}
    \label{bin}
\end{figure}

\par In this section, we investigate the evolution of the binary fraction in our models, defining the binary fraction as bf=bs/(bs+ss), where "bs" is the number of binary systems and "ss" is the number of single stars. Figure~ \ref{bin} shows the evolution of the binary fraction in all models. Panel (a) presents the fraction of binaries, and panel (b) shows the evolution of the fraction of observable binaries compared to the observational data over time. We emphasize that these two quantities are calculated from different stellar populations. The binary fraction shown in panel (a) is measured for the entire simulated cluster, including all stellar masses and magnitudes, whereas the observable binary fraction in panel (b) is determined only from the stars that satisfy the \textit{Gaia} mass and magnitude detection limits.

\par The initial presence of substructure is a key driver in the dynamical evolution of binary stars within clusters (\citealt{kupper, parker14, daff, tor} and \citealt{amiri}). The sharp initial decrease in the binary fraction during the first $\sim 2$ Myr, observed in all models, can be attributed to the rapid dynamical processing of primordial binaries within the initial substructures. Since all models contain a degree of initial substructure, binaries experience more frequent stellar encounters in locally dense regions, where the dynamical timescales are considerably shorter than the global cluster timescale (\citealt{parker14} and \citealt{dorval17}). These encounters preferentially disrupt weakly bound binaries (\citealt{pang2026}), including systems that may be dynamically soft despite the adopted maximum semi-major axis of 50 AU. The initial decrease is stronger in models with a high degree of initial substructure and rotation. Stronger rotation slows the rapid merging of dense subregions and prolongs their survival. After this early phase, the number of vulnerable binaries is reduced, leaving a population that is generally more resistant to further disruption. Consequently, the binary fraction decreases more slowly, although dynamical interactions continue within the surviving substructures.

\par After approximately 50 Myr, the dominant mechanism driving binary disruption shifts in models with significant initial rotation. Rotation itself becomes the primary factor governing the continued decline in the binary fraction, consistent with findings by \citealt{kamlahrot} and \citealt{pang2025}. The late-time evolution of the binary fraction is determined by the combined effects of internal binary disruption and processing through dynamical encounters, together with the gradual depletion of the cluster caused by rotation-enhanced mass loss.

\par Panel (b) shows that the observable binary fractions of all models remain consistent with the values inferred for observed star clusters. We note that these fractions are systematically higher than the intrinsic primordial binary fractions of 10-25\% listed in Table \ref{table}. This difference results from the Gaia limits, which preferentially exclude faint, low-mass stars and therefore bias the observable sample toward higher-mass stars, where the binary fraction is expected to be higher (\citealt{sana, oh15} and \citealt{moe2017}). This agreement supports the choice of a primordial binary fraction of 10-25\% as a more realistic initial condition. 

\par It is important to distinguish between two different effects contributing to the trends shown in Fig.~\ref{bin}: (i) the continuous decrease of the binary fraction in panel (a) indicates that the disruption of primordial binaries dominates over the formation of new binaries during the entire evolution. (ii) in panel (b), in contrast, the late-time ($\geq$ 500 Myr) increase in the observable binary fraction does not imply binary formation, but rather results from observational selection effects: preferential loss of low-mass stars through evaporation and rotation-enhanced mass loss, combined with mass segregation, increases the relative contribution of higher-mass stars with intrinsically larger binary fractions. Together, these effects explain the difference between the trends observed in the two panels.

\subsection{Evolution of angular momentum}
\label{momentumsec}
\begin{figure}[htb]
   \centering
\includegraphics[scale=0.38]{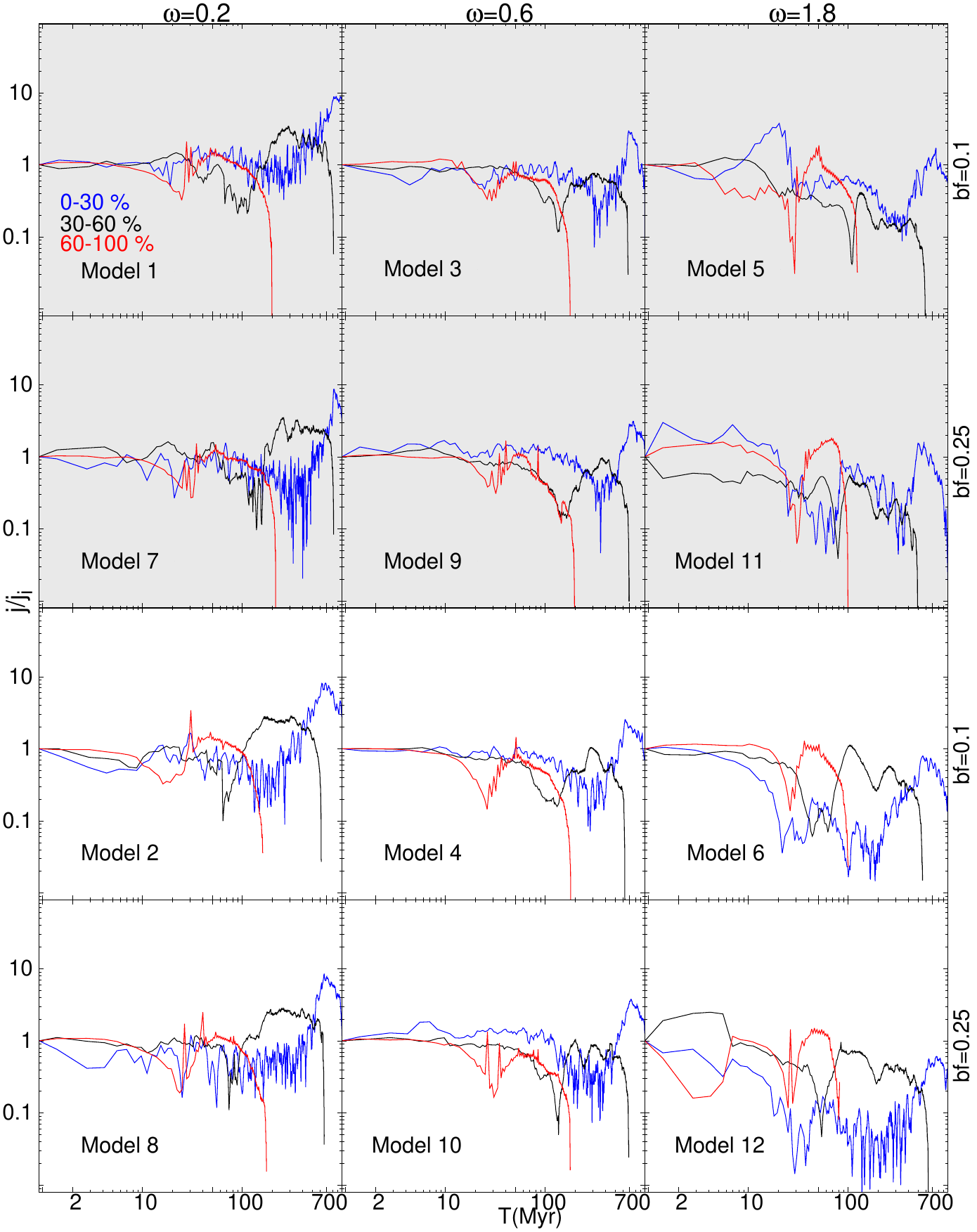}

    \caption{Evolution of j$/$j$_i$ for different Lagrangian radii for all models. Panel distribution represents different models, as in Fig.~ \ref{fr}. }
    \label{jtlagr}
\end{figure}

\begin{figure}[htb]
   \centering
\includegraphics[scale=0.37]{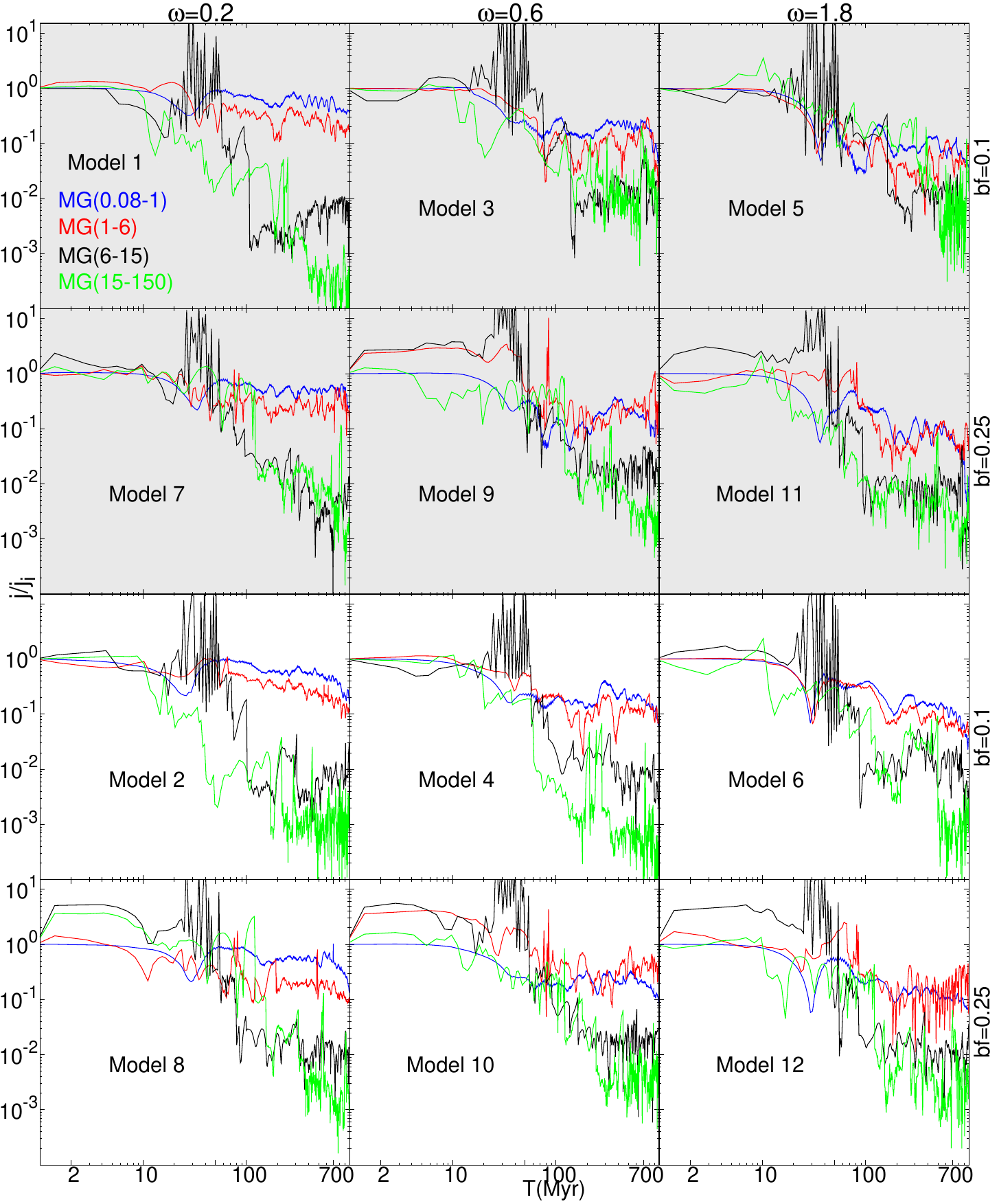}
 
    \caption{Evolution of j$/$j$_i$ for different mass groups of stars for all models. Panel distribution represents different models, as in Fig.~ \ref{fr}.  }
    \label{jtm}
\end{figure}

\par A key element in understanding the dynamical evolution of our models is the evolution of the normalized specific angular momentum, j$/$j$_i$. For each selected group of stars, we calculate
j $= \left| \sum_{k \in N_{s,t}} \mathbf{r}_k \times \mathbf{v}_k \right|,$
where $N_{s,t}$ denotes the stars belonging to the considered group at time $t$ (here, j$_i$ is j at $t=$0 ).

To analyze this process, Fig.~ \ref{jtlagr} shows the temporal evolution of j$/$j$_i$ in three Lagrangian ( inner: 0-30 \%, middle: 30-60 \%, and outer: 60-100 \% ) regions with respect to the initial total mass (as in Sect.~\ref{lagrsec}), while Fig.~\ref{jtm} presents j$/$j$_i$, broken down by zero-age main sequence (ZAMS) stellar mass ranges. For clarity, we define four ZAMS mass groups (in $M_\odot$): very low mass MG(0.08–1), low mass MG(1–6), high mass MG(6–15), and very massive MG(15–150).

\par As illustrated in Fig.~ \ref{jtlagr}, during the early evolutionary phase ($t \sim $ 10-50 Myr), j$/$j$_i$ of the outer region initially decreases and subsequently increases in all models. The onset of this decline approximately coincides with the beginning of stellar escape (Fig.~ \ref{mass}). Since escaping stars are preferentially removed from the outer regions, they carry away angular momentum and reduce the angular momentum of these regions. Subsequently, the angular momentum of the inner and middle regions decreases, while that of the outer region increases, indicating an outward redistribution of angular momentum. This transfer is more evident in models with the high initial rotation, producing a sharp increase in the outer region and a corresponding decline in the inner and middle regions. After approximately 100 Myr, the middle region undergoes a similar decrease followed by an increase. This early redistribution of angular momentum is strongest in models with the higher initial rotation and a lower degree of initial substructures, consistent with the findings of \citealt{spurzem99rot, kim2004}, and \citealt{pang2025}. Conversely, models with a high degree of initial substructures or a larger initial binary fraction exhibit weaker redistribution and a more modest early exchange of angular momentum between the inner and outer regions. In highly substructured models, the rapid merging of substructures produces a violent early evolution that modifies this otherwise regular behavior.

\par After approximately $t \sim 50$ Myr, the evolution of j$/$j$_i$ changes qualitatively. The outer region undergoes a pronounced decrease in j$/$j$_i$, with the decline occurring more rapidly in models with stronger initial rotation. Meanwhile, j$/$j$_i$ in the middle region increases before declining at later times, when the inner region begins to increase. This radial sequence indicates a gradual inward redistribution of angular momentum. The sharp decline or truncation of the outer-region curves is mainly caused by tidal escape and the strong depletion of bound stars, which remove a substantial amount of angular momentum from the outer regions. At later times ($>$500 Myr), a similar trend appears in the next outer region, suggesting that angular momentum is transported inward in a stepwise manner. The late-time increase in the angular momentum of the innermost regions is consistent with the results of \citealt{tiongco2017}, who reported that the inward shift of the rotation peak occurs only during advanced evolutionary stages. At these late times, our models reach comparable dynamical ages, allowing angular momentum to become increasingly concentrated toward the inner regions.

\par As Fig.~ \ref{jtlagr} shows, all models exhibit an overall decline in angular momentum during the simulations in their outer and middle regions, in agreement with previous studies (\citealt{spurzem99rot, kim2008, tiongco2018} and \citealt{pang2025}). However, the inner regions do not follow the same monotonic decline. Instead, their angular momentum increases at later evolutionary times in models with the low and moderate initial rotation, as well as in models combining high initial rotation with a high degree of initial substructures. This differs from the results of \citealt{pang2025}, who reported a strong decline in the angular momentum of the inner regions in models with the high initial rotation. In our simulations, models with the low and moderate initial rotation, all of which contain some degree of initial substructures, show a substantial late-time increase in the angular momentum of the inner region. A similar increase occurs in models combining high initial rotation with a high degree of initial substructures. This discrepancy may result from the combined effects of initial substructures and the different stellar populations considered, as \citealt{pang2025} focused on a restricted stellar mass range, whereas our analysis includes all stars in the cluster. We also note that the number of particles in the inner regions becomes considerably smaller at these late evolutionary times, which may affect the measured trend. Nevertheless, our results are consistent with \citealt{pang2025} for models combining high initial rotation with a low degree of primordial substructure. The overall decline of j$/$j$_i$ shown in Fig.~ \ref{jtlagr} is also qualitatively consistent with the recent observations of \citealt{dalessandro2026}, who found that 50-60\% of clusters younger than 500 Myr exhibit significant rotation, compared with only $\sim$15\% of older clusters.

\par Figure~ \ref{jtm} shows that, overall, j$/$j$_i$ decreases with time for all stellar mass groups. This global decline reflects the gradual redistribution and loss of angular momentum as the clusters evolve (\citealt{kim2004} and \citealt{kamlahrot}). However, the exact evolution of j$/$j$_i$ is strongly influenced by stellar mass, the initial rotation of the cluster, its initial substructure, and the primordial binary fraction.

\par In the early stages of evolution ( $\sim$ 20-50 Myr), all models exhibit pronounced fluctuations in j$/$j$_i$ for the  group MG$(6$–$15)$. These fluctuations are followed by a sharp decline. This behavior is characteristic of strong mass loss and, therefore, loss of angular momentum by stellar evolution in competition with mass segregation. We note that the fluctuations observed between $\sim$20–50 Myr in MG(6–15) and in the outer Lagrangian shell (Fig.~ \ref{jtlagr}) likely originate from related physical processes. During this phase, a significant fraction of stars in this mass group reach the end of their main-sequence lifetimes and undergo substantial mass loss through stellar winds and supernova explosions. This leads to the removal of both mass and angular momentum from the cluster. At the same time, transient enhancements and fluctuations in the mass segregation signal become more prominent due to rapid dynamical rearrangement of massive stars. These combined effects increase the exchange of stars across the outer Lagrangian boundary, which results in the observed fluctuations.  After these fluctuations, j$/$j$_i$ decreases substantially for MG(6-15) and MG(15-150), whereas it remains comparatively stable for MG(0.08-1) and MG(1-6). Given the increasing number of escapers in all models, the weaker decline of the lower mass groups indicates that their angular momentum loss is partially compensated by angular momentum redistribution from more massive stars, consistent with the results of \citealt{kim2004} and \citealt{kamlahrot}.

\par After $\sim100$ Myr, a clear mass-dependent separation emerges in the evolution of j$/$j$_i$ which agrees with the observations reported by \citealt{scalco2023}, who found that dynamically evolved clusters exhibit a correlation between rotation and stellar mass. The two groups, MG(6-15) and MG(15-150), experience a significantly stronger decline in j$/$j$_i$ as these massive stars have evolved and lost mass and angular momentum through stellar winds and supernova explosions by that time. As the initial rotation increases, this mass-dependent separation becomes weaker, indicating that stronger initial rotation promotes a more efficient redistribution of angular momentum across different mass groups and reduces the contrast between them. In the time interval between $\sim$ 100 and 700 Myr, models with a high degree of initial rotation and a low fraction of primordial binaries show a noticeable increase in j$/$j$_i$ for the group MG(15-150).

\par An additional noteworthy feature appears in the models with a low degree of initial substructures and a high fraction of initial binary systems. In these models, when the initial rotation is high, MG(1-6) shows a pronounced increase in j$/$j$_i$. The smoother initial structure suppresses early chaotic angular-momentum loss, while the high binary fraction enables efficient angular-momentum exchange.

\subsection{Changing in shape of the clusters}
\begin{figure}[htb]
   \centering
\includegraphics[scale=0.38]{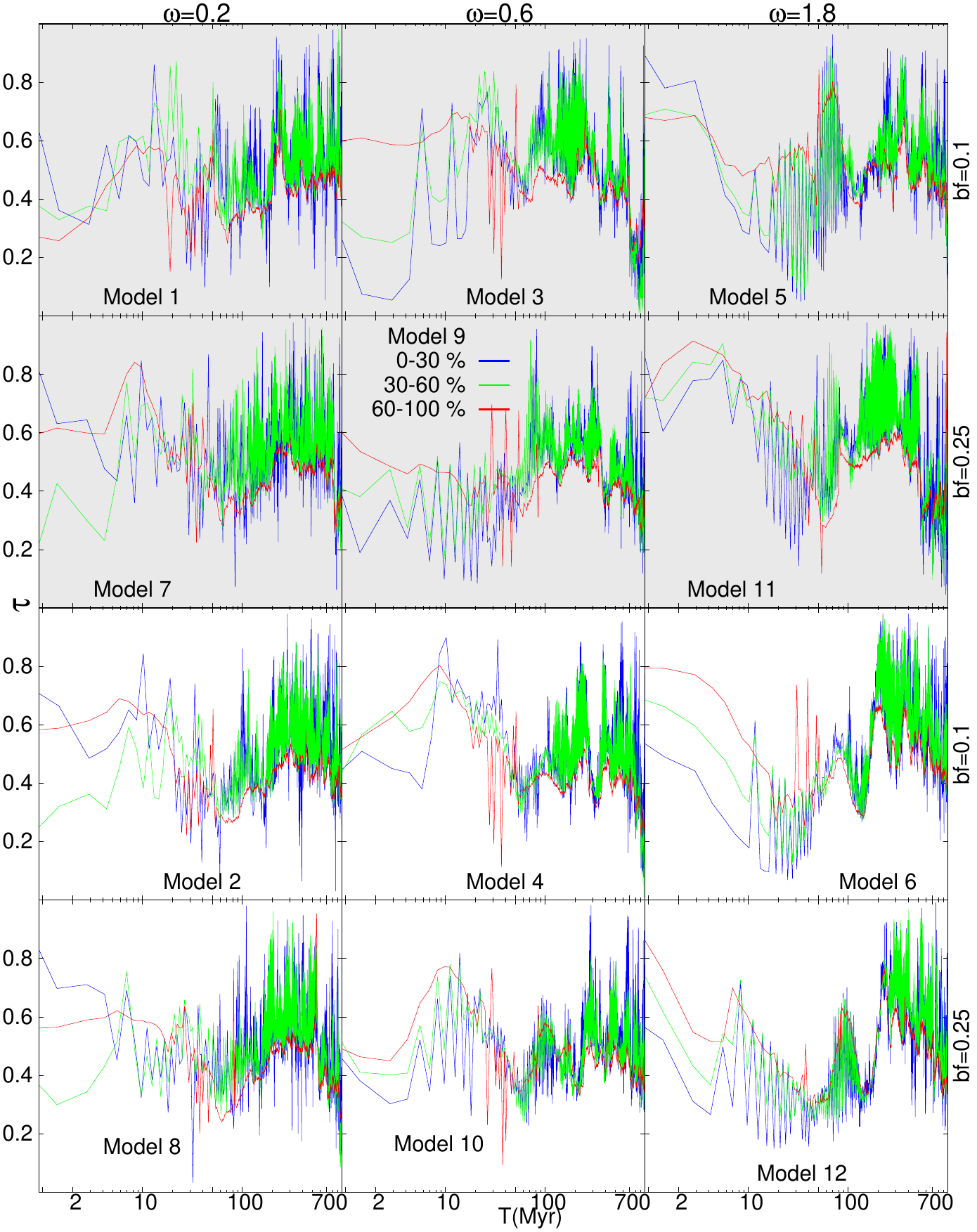}
 
    \caption{Evolution of triaxiality for all models. Panel distribution represents different models, as in Fig.~ \ref{fr}.}
    \label{tr}
\end{figure}

\begin{figure}[htb]
   \centering
\includegraphics[height=60mm, width=85mm]{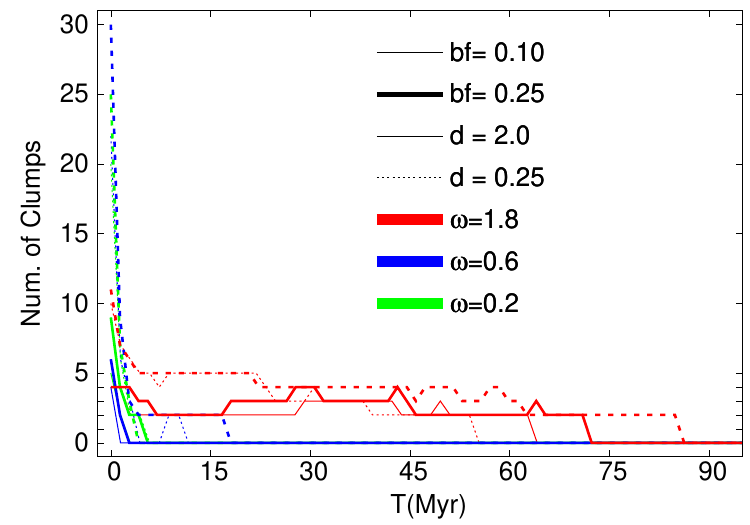}
 
    \caption{Time evolution of the number of clumps for all models. The models are shown using the same line styles and colors as in Fig.~ \ref{lagr}.}
    \label{nclumps}
\end{figure}

\begin{figure}[htb]
   \centering
\includegraphics[scale=0.4]{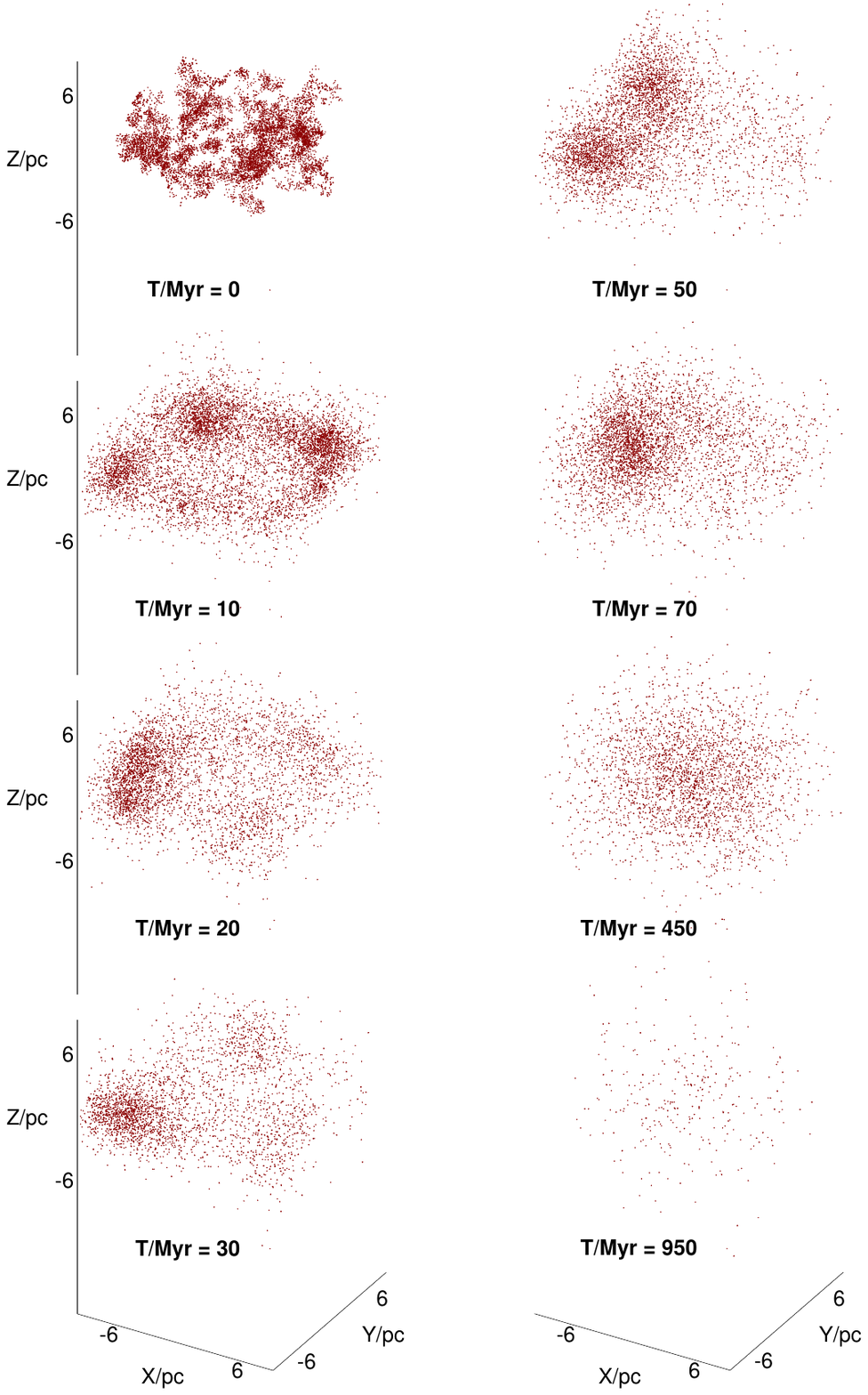}
 
    \caption{Snapshots from eight different steps of the evolution of Model 11.}

    \label{ev3d}
\end{figure}

\par Studying the parameter $\tau$, which is defined as $\tau=\frac{a^2 - b^2}{a^2 - c^2}$ in which $a>b>c$ are the semi-axes of an ellipsoid, is one way to examine the shape of the clusters as they evolve.  According to this description, a cluster with $\tau$ values close to 1 is almost bar-shaped and shows prolate structure, but a cluster with values close to 0 has an oblate shape, resembling a disk.

\par To investigate the triaxiality of the systems, we emphasize that our models simultaneously include three key ingredients: a degree of initial substructures and rotation, and a fraction of binary systems. The coexistence of these features implies that the cluster shape is expected to evolve through a combination of different physical mechanisms, including rotational flattening driven by centrifugal forces, the merging of initial substructures, and dynamical heating produced by binary interactions.

\par Figure~ \ref{tr} shows the time evolution of the triaxiality parameter $\tau$ for all models. The quantity $\tau$ is computed separately for three Lagrangian ( inner: 0-30 \%, middle: 30-60 \%, and outer: 60-100 \% ) regions. We note that at late evolutionary stages, when a substantial fraction of stars has already escaped, the measured values of $\tau$ become less reliable, particularly in the inner region.

\par For models with a low degree of initial rotation, $\tau$ exhibits large and rapid fluctuations throughout the evolution. This behavior reflects the highly time-dependent and irregular evolution of the cluster shape in these systems. Clearer trends emerge, however, as the degree of initial rotation increases. In models with the low initial substructures and moderate to high degrees of rotation, the inner Lagrangian region shows a tendency toward a disk-like configuration before $\sim100$ Myr. This signal becomes more pronounced as the rotation parameter increases, even for the middle and outer regions, in agreement with reported results from \citealt{pang2025}. In contrast, for models with a high degree of initial substructures, the same disk-like signature is noticeably weaker, indicating that substructure partially suppresses the flattening induced by rotation.

\par At later times, increasing the degree of initial rotation leads to the emergence of a bar-like shape, consistent with results from \citealt{kamlahrot} and \citealt{pang2025}, most clearly seen in the inner regions during the interval $\sim$ 100-700 Myr. The strength of this bar-shaped signature depends on the initial conditions. In particular, it becomes weaker in models with a high degree of initial substructures and a low initial binary fraction. Note that, toward the end of this interval ($\gtrsim$ 500 Myr), the decreasing number of bound stars leads to increasing uncertainties, making the measured $\tau$ values noisier.

\par To quantify how the different initial configurations influence the formation and survival of such subclusters, we compute the number of clumps as a function of time for all models, using the \textsc{Hierarchical Density-Based Spatial Clustering of Applications with Noise} (\textsc{HDBSCAN} \footnote{The HDBSCAN parameters are set as follows: \textsc{min-cluster-size=max(100,0.01N)}, where N is the number of stars in each snapshot, and \textsc{cluster-selection-method=leaf}. Candidate clumps are retained only if their local density exceeds 10\% of the global density.}, \citealt{hdbscan}) that has been used for galaxy classification (\citealt{logan2020}) and the merger history of the Milky Way (\citealt{sante2026}). 

\par As shown in Fig. \ref{nclumps}, all models initially contain a substantial number of subclumps as a consequence of the initial substructure. During the subsequent dynamical evolution, models with the higher rotational support retain a larger number of clumps for a longer duration. This behavior indicates that rotation acts to stabilize the substructure against rapid dissolution by reducing the efficiency of early violent relaxation.

\par The results show that the combination of high rotation and strong initial substructures is the dominant factor in sustaining subclusters during the first $\sim20$ Myr. At later times, the role of the primordial binary population becomes increasingly important: the higher initial binary fraction contributes additional dynamical heating that helps preserve small-scale substructures, allowing clusters to maintain two to four subclusters up to $\sim70$ Myr.

\par About the global evolution of the cluster morphology, Fig.~ \ref{ev3d} illustrates the visual structural changes in several snapshots of the model characterized by the highest degree of rotation and initial substructures. This figure demonstrates that strong rotation promotes the formation of a persistent group (or binary) of subclusters within the main system. These subclusters remain gravitationally bound for an extended period. Since observations report the existence of binary and group star clusters (\citealt{palma2025}), 
our results provide insights into the survival of compact internal subclusters, emerging from the effects of a high degree of initial rotation and substructure combined with a high fraction of primordial binary systems. It is important to distinguish these internal subclusters from the binary and grouped open clusters reported by \citealt{palma2025}, where the individual clusters are separate gravitationally bound systems with separations of up to 50 pc and many systems have ages younger than 100 Myr. In contrast, our simulated substructures are more compact components within a single cluster that eventually merge or dissolve. Nevertheless, their long-term survival may provide insight into early hierarchical assembly processes. For example, \citealt{della2025} found that the \textsc{LISCA I} and \textsc{W345} systems formed approximately 20–30 Myr ago, with components separated by a few hundred parsecs. While our simulations do not directly reproduce such large-scale systems, the survival of substructures for up to $\sim$70 Myr suggests that rotation and primordial substructure can influence the early evolution of clustered environments.

\section{Conclusions}
\label{conc}

In this paper, we simulated the evolution of twelve cluster models for one Gyr using the \textsc{Nbody6++GPU} code. For the first time, these simulations accounted for varying degrees of initial substructures, primordial binary fractions, and bulk rotation. This comprehensive approach enabled us to explore more realistic initial conditions for star clusters. The main results are summarized as follows:

\begin{itemize}

\item[1-] We developed a method to generate initial conditions that consistently incorporate both initial substructure and bulk rotation within the Fokker–Planck framework. This method produces clusters with realistic substructures while preserving their global rotational properties, naturally reproducing the flattened morphology expected for rotating stellar systems.
        
\item[2-] The lifetime of substructures strongly depends on the degree of rotation. High initial rotation generates significant centrifugal support, slowing the collapse and merging of substructures and extending their survival. Our results show that moderate rotation helps preserve initial substructures for longer timescales in young clusters, consistent with observations. The Q parameter also depends on the radius used for its calculation: larger volumes systematically produce higher Q values, reflecting a smoother large-scale stellar distribution compared to inner regions. In models with the high initial rotation, inner regions retain substructures for almost the entire simulation. For low to moderate rotation, the inner region shows two phases: substructures disappear on a timescale similar to that in the outer regions, but the Q parameter rapidly decreases. Although this decline resembles the signature of a highly substructured distribution, it is consistent with changes in the central density distribution associated with core contraction and mass segregation. This behavior occurs earlier as the initial rotation becomes weaker.

\item[3-] In the first $\sim 10$ Myr, clusters with the low rotation and a high initial degree of substructure show a clear contraction in $r_{10}$, $r_{30}$, and $r_{50}$. During this early phase, the average stellar mass inside the inner Lagrangian radii is systematically higher in models with the high initial substructure and low to moderate rotation. In models with the high initial rotation, the increase in central average mass appears later, especially when the primordial binary fraction is low. The contraction before $\sim 60$ Myr in highly rotating models is consistent with gravothermal-gravogyro evolution, and it appears earlier in the very inner regions when the degree of initial substructures is high and the primordial binary fraction is low.

\item[4-] After the substructures merge, the clusters gradually develop clearer and more persistent signatures of mass segregation. Two processes dominate the early evolution: rotational centrifugal forces, which delay the emergence of mass segregation, and the merging of initial substructures, which enhances it. In clusters with a low degree of initial substructures, a high initial binary fraction also leads to rapid signatures of mass segregation.

\item[5-] As reported in previous studies, initial substructures increase mass loss in non-rotating clusters. However, in rotating clusters, a high degree of initial substructures leads to milder early dynamics and suppresses early escapers. Therefore, the mass-loss behavior of substructured clusters depends critically on the interplay between substructure and rotation.

\item[6-] Even with a high degree of initial substructures, initial rotation strongly influences the long-term evolution of the binary population. However, the late-time decline in the binary fraction results from the combined effects of internal dynamical processing through binary encounters and the enhanced global mass loss associated with rotating clusters.

\item[7-] We note an important difference from the results of \citealt{pang2025}, who reported a strong decrease in angular momentum in the inner regions for models with the high initial rotation. In contrast, our models with a high degree of initial substructures show a substantial increase in the angular momentum of the innermost regions at late times, even when the initial rotation is strong.

\item[8-] In models with the low initial substructures and moderate to high rotation, the inner Lagrangian region tends to form a disk-like configuration before $\sim100$ Myr, with this signature becoming stronger as the rotation parameter increases, even in middle and outer regions. For models with the high initial substructures, this disk-like feature is weaker, indicating that substructure partially suppresses rotation-induced flattening. At later times, higher initial rotation leads to a bar-like shape, most evident in the inner regions between $\sim$100–700 Myr, but this feature is weaker in models with the high initial substructures and low initial binary fraction. However, the apparent persistence of the bar-like shape after $\sim$ 500 Myr remains tentative, as the decreasing number of bound stars increases uncertainties in the shape measurements.

\item[9-] This study shows that high initial rotation, combined with a high degree of initial substructures and a large fraction of primordial binary systems, promotes the survival of internal subclusters within the main cluster, thereby providing insights into the early survival and dynamical evolution of hierarchical stellar substructures.

\end{itemize}

 The cluster masses reported by \citealt{pang22} and \citealt{pang23} are present-day masses shaped by their individual dynamical evolution. In contrast, our models cover only two initial masses and a limited range of rotation, substructure, and binary fraction. This difference may partly explain the scatter of the observations around our models in Figs.~\ref{fr}, \ref{seg}, \ref{mass} and \ref{bin}. In addition, as we performed only one simulation for each model, the sensitivity of individual evolutionary features to different random realizations of the same initial configuration remains unexplored. For future work, it would be valuable to extend these simulations to larger particle numbers and to explore different tidal field models, in order to obtain a more comprehensive understanding of cluster evolution in the presence of these features. We emphasize that the comparisons with observational data are intended to assess the qualitative consistency of the simulated evolutionary trends. Our twelve models cover only a coarse grid of initial rotation, substructure, and binary fraction, representing low, moderate, and high regimes. Nevertheless, the comparisons indicate that models with the high initial substructure, moderate to high initial rotation, and relatively high primordial binary fractions tend to agree more closely with the observed trends. This should be viewed only as an indication of the parameter-space region for further studies.

\begin{acknowledgements}

We thank the anonymous referee for the constructive comments and valuable suggestions. VA acknowledges DAAD for supporting this study to be conducted (Funding Program ID: 57588370). VA would like to thank Kai Wu for the useful discussions and constructive comments during the preparation of this paper. VA and RS also acknowledge NAOC International Cooperation Office for its support in 2023, 2024, and 2025; RS acknowledges Chinese Academy of Sciences President's International Fellowship Initiative for Visiting Scientists (PIFI, grant No. 2026PVA0089), the National Natural Science Foundation of China (NSFC) under grant No. 12473017, and the German Science Foundation DFG (Grant Sp 345/24-1). 
This material is based upon work supported by Tamkeen under the NYU Abu Dhabi Research Institute grant CASS.

PB appreciates the Polish high-performance computing infrastructure PLGrid (HPC Centre: ACK Cyfronet AGH) for providing computer facilities and support within computational grant No. PLG/2026/019243. 
X.P. acknowledges the financial support of the National Natural Science Foundation of China through grants 12573036 and 12233013, and the China Manned Space Program with grant No. CMS-CSST-2025-A08.
MAS acknowledges funding from the European Union’s Horizon 2020 research and innovation programme under the Marie Skłodowska-Curie grant agreement No.~101025436 (project GRACE-BH) and from the MERAC Foundation through the 2023 MERAC prize. MAS acknowledges the ACME project, which has received funding from the European Union's Horizon Europe Research and Innovation programme under Grant Agreement No. 101131928.
\end{acknowledgements}

\bibliographystyle{aa}
\bibliography{UCDs}
\end{document}